\documentclass[journal,twoside,web]{ieeecolor}
\usepackage{generic}
\usepackage{cite}
\usepackage{amsmath,amssymb,amsfonts}
\usepackage{algorithmic}
\usepackage{graphicx}
\usepackage{algorithm,algorithmic}
\usepackage{hyperref}
\hypersetup{hidelinks=true}
\usepackage{textcomp}

\graphicspath{{figures/}}
\usepackage{subcaption}
\usepackage{tikz}
\usepackage{tikzpeople}
\usetikzlibrary{calc}
\usepackage[dvipsnames]{xcolor}
\usepackage{siunitx}
\usepackage{xspace}
\usepackage{comment}

\begin{document}
\title{Neural Tracking of Sustained Attention, Attention Switching, and Natural Conversation in Audiovisual Environments using Mobile EEG}
\author{Johanna Wilroth\textsuperscript{$\star$}\textsuperscript{\textdagger}, Oskar Keding\textsuperscript{$\star$}\textsuperscript{\textdagger}, Martin A. Skoglund \IEEEmembership{Member, IEEE}, \\ Maria Sandsten, Martin Enqvist and Emina Alickovic, \IEEEmembership{Member, IEEE}
\thanks{\textsuperscript{$\star$} Equally contributed as first authors.}
\thanks{\textsuperscript{\textdagger} Authors to whom any correspondence should be addressed.}
\thanks{Submitted for review on 2025-12-05. This work was supported by the ELLIIT Strategic Research Programme. We thank all participants for taking part in this study and the research audiologists at Eriksholm Research Centre for their assistance with data collection.}
\thanks{Johanna Wilroth is with the Dept. of Electrical Engineering, Linköping University, Sweden (johanna.wilroth@liu.se).}
\thanks{Oskar Keding is with the Centre for Mathematical Sciences, Lund University, Sweden (oskar.keding@matstat.lu.se).}
\thanks{Martin A. Skoglund is with the Dept. of Electrical Engineering, Linköping University, Sweden and Eriksholm Research Centre, Oticon A/S, Denmark (mnsk@eriksholm.com).}
\thanks{Maria Sandsten is with the Centre for Mathematical Sciences, Lund University, Sweden (maria.sandsten@matstat.lu.se).}
\thanks{Martin Enqvist is with the Dept. of Electrical Engineering, Linköping University, Sweden (martin.enqvist@liu.se).}
\thanks{Emina Alickovic is with the Dept. of Electrical Engineering, Linköping University, Sweden and Eriksholm Research Centre, Oticon A/S, Denmark (eali@eriksholm.com).}
}

\maketitle

\begin{abstract}
Everyday communication is dynamic and multisensory, often involving shifting attention, overlapping speech and visual cues. Yet, most neural attention tracking studies are still limited to highly controlled lab settings, using clean, often audio-only stimuli and requiring sustained attention to a single talker.  This work addresses that gap by introducing a novel dataset from 24 normal-hearing participants. We used a mobile electroencephalography (EEG) system (44 scalp electrodes and 20 cEEGrid electrodes) in an audiovisual (AV) paradigm with three conditions: sustained attention to a single talker in a two-talker environments, attention switching between two talkers, and unscripted two-talker conversations with a competing single talker. Analysis included temporal response functions (TRFs) modeling, optimal lag analysis, selective attention classification with decision windows ranging from 1.1$\,\text{s}$ to 35$\,\text{s}$, and comparisons of TRFs for attention to AV conversations versus side audio-only talkers. Key findings show significant differences in the attention-related P2-peak between attended and ignored speech across conditions for scalp EEG. No significant change in performance between switching and sustained attention suggests robustness for attention switches. Optimal lag analysis revealed narrower peak for conversation compared to single-talker AV stimuli, reflecting the additional complexity of multi-talker processing. Classification of selective attention was consistently above chance (55–70\% accuracy) for scalp EEG, while cEEGrid data yielded lower correlations, highlighting the need for further methodological improvements. These results demonstrate that mobile EEG can reliably track selective attention in dynamic, multisensory listening scenarios and provide guidance for designing future AV paradigms and real-world attention tracking applications. 
\end{abstract}


\def\tvpersonleft#1#2#3{
\begin{scope}[shift={#1}, rotate=#2]
    \clip (-1,-0.6) rectangle (1,0.8);
    \node[#3,minimum size=1.2cm,rotate=#2]{};
    \draw (-1,-0.6) rectangle (1,0.8);
\end{scope}
}

\def\tvpersonright#1#2#3{
\begin{scope}[shift={#1}, rotate=#2]
    \clip (-1,-0.6) rectangle (1,0.8);
    \node[#3,mirrored,minimum size=1.2cm,rotate=#2]{};
    \draw (-1,-0.6) rectangle (1,0.8);
\end{scope}
}

\def\speaker#1#2#3#4{
\begin{scope}[shift={#1}, rotate=#2]
    \draw[#4,thick] (0*#3,0.5*#3)--(0.5*#3,1*#3)--(0.5*#3,-1*#3)--(0*#3,-0.5*#3)--(0*#3,0.5*#3)--(-0.25*#3,0.5*#3)--(-0.25*#3,-0.5*#3)--(0*#3,-0.5*#3);
\end{scope}
}

\def\sustname{sustained attention condition\xspace}
\def\sustnamemid{sustained attention\xspace}
\def\sustnameshort{\mbox{SustAC}\xspace}

\def\switname{switching attention condition\xspace}
\def\switnamemid{switching attention\xspace}
\def\switnameshort{\mbox{SwitAC}\xspace}

\def\convname{conversation attention condition\xspace}
\def\convnamemid{conversation attention\xspace}
\def\convnameshort{\mbox{ConvAC}\xspace}

\definecolor{pyblue}{rgb}{0.121, 0.466, 0.705}
\definecolor{pyorange}{rgb}{1.000, 0.498, 0.055}
\definecolor{pygreen}{rgb}{0.173, 0.627, 0.173}
\definecolor{pyred}{rgb}{0.839, 0.153, 0.157}
\definecolor{johanna}{RGB}{150,50,200}
\newcommand{\johanna}[1]{\textcolor{johanna}{\textbf{#1}}}
\newcommand{\oskar}[1]{\textcolor{pyred}{\textbf{#1}}}

\section{Introduction}
\label{sec:int}

Neurophysiological studies of speech communication have traditionally relied on highly controlled stimuli-—short, repetitive sounds or isolated sentences—-designed to obtain time-locked evoked responses such as event related potentials (ERPs) \cite{stapells2002cortical}. While these controlled paradigms allow for well-defined experimental contrasts, they fail to capture the neural dynamics involved in real-world listening. A recent shift has occurred toward using more naturalistic stimuli,  largely driven by the development of data-driven models to decode cognitive states such as auditory attention. This transition reflects a growing ambition to understand and track selective auditory attention in realistic, everyday listening situations, an essential step toward building EEG-based auditory attention decoding (AAD) systems that can infer the focus of a listener’s attention \cite{o2015attentional,alickovic2019tutorial,geirnaert2021electroencephalography} for future hearing devices \cite{lunner2020three, di_liberto_speech_2025}.

EEG has become the primary neuroimaging modality for AAD research due to its portability and compatibility with wearable systems \cite{debener2015unobtrusive, kappel2018dry}. Yet, despite this potential, current electroencephalography (EEG) based AAD paradigms remain far from reflecting everyday listening.  
Most studies remain confined to laboratory environments, using clean, often audio-only, stimuli and requiring sustained attention to a single talker for extended periods. Such designs contrast sharply with natural communication, which is multisensory, noisy, and often requires switching attention between speakers \cite{keidser2020quest, bodie2023listening}. These  simplifications limit both the ecological validity and practical relevance for integration into hearing technologies.

The path toward tracking attention in naturalistic settings faces three critical roadblocks. 
First, there is an over-reliance on clean, controlled, and often audio-only speech stimuli \cite{o2015attentional, schafer2018testing, ciccarelli2019comparison, alickovic2019tutorial, alickovic2021effects, geirnaert2021electroencephalography, straetmans_neural_2024}. Most studies use speech recorded by professional speakers, with controlled pauses and minimal background noise, which does not reflect the complexity of real-world communication. Everyday communication is inherently multisensory \cite{Kriegstein2021}, often accompanied by visual cues such as lip movements and facial expressions. Recognizing this, recent efforts have started to introduce audio-visual (AV) stimuli into AAD experiments, showing that visual speech cues can enhance neural tracking and decoding of auditory attention \cite{crosse2015congruent, o2019look, fu2019congruent, wang2023eeg, rotaru2024we}. However, these AV paradigms still frequently rely on scripted speech and controlled interactions, which differ from the spontaneous, overlapping, and noisy sound flow of everyday communication. To bridge this gap, AAD paradigms should integrate naturalistic AV speech that captures the dynamic and unscripted nature of real-life interactions.

Second, there is a limited understanding of how neural speech processing adapts across different listening tasks. Traditional AAD paradigms, which require participants to sustain attention on a single (usually audio-only) talker for prolonged durations\cite{o2015attentional, bleichner_identifying_2016, schafer2018testing, alickovic2019tutorial, geirnaert2021electroencephalography, belo2021eeg}, fail to capture the dynamic shifts in attention characteristic of real-world listening. Recognizing this gap, recent studies have begun exploring attention-switching paradigms  \cite{haro2022eeg, van_de_ryck_eeg-based_2025,carta_simultaneous_2025}. For example, Haro et al. \cite{haro2022eeg}  showed that while attention switches between two audio-only speakers can be decoded, these switches are associated with increased listening effort. Carta et al. \cite{carta_simultaneous_2025} further examined neural dynamics during attention shifts, observing that neural tracking of a newly attended speaker emerges even before disengagement from the previous one, suggesting a brief period of simultaneous encoding of both streams. Extending these investigations to AV contexts, Van de Ryck et al. \cite{van_de_ryck_eeg-based_2025} demonstrated that AAD performance during switching trials is comparable to sustained attention trials in multi-conversation audiovisual settings. However, these studies either focus on neural dynamics or on decoding attention, but not both together during attention switches in complex AV conversations. This leaves a gap in understanding how attention decoding performs when moving from simple two-talker tasks to more natural, multi-talker conversations. 

Third, achieving realistic AAD requires more compact and less intrusive EEG systems. 
Mobile EEG systems, capable of recording neural activity using either full scalp electrode arrays or more discreet in/around-ear electrodes (e.g., in-ear EEG \cite{kappel2018dry}  or cEEGrid \cite{debener2015unobtrusive} arrays), offer a promising solution for capturing neural responses in ecologically valid environments.
 
The cEEGrid, a flexible array of electrodes placed around the ear, has been validated to capture auditory attention signatures. 
However, its application has predominantly been limited to simplified experimental conditions, such as sustained attention to a single talker over extended durations  \cite{bleichner_identifying_2016,holtze_ear-eeg_2022,mirkovic_target_2016}. 
Additionally, current cEEGrid-based AAD approaches often rely on individual parameter tuning, which may introduce subject-specific biases in performance evaluation. Recent advancements, such as the work by Straetmans et al. \cite{straetmans_neural_2024}, have begun addressing these limitations by leveraging portable mBTrain amplifiers in combination with cEEGrid arrays to conduct AAD experiments outside the laboratory. However, results based on cEEGrid electrode measurements in these complex settings are yet to be reported.

To address these three roadblocks, we recorded EEG from 24 normal-hearing participants using 44 scalp electrodes and cEEGrid arrays across three conditions: sustained attention to one of two competing talkers, attention switching between two competing talkers, and attending to a conversational AV source with a competing single talker. To overcome the first roadblock of limited ecological validity, we used naturalistic audiovisual stimuli. To address the second roadblock of dynamic attention demands, we included both sustained and switching attention tasks, as well as attention to conversation, each with competing, ignored speech present. To tackle the third roadblock of limited portability and practical EEG application, we employed mobile EEG recording with both scalp and around-the-ear cEEGrid electrodes. Notably, the study was conducted in Danish, a language not previously examined in this context, enabling assessment of potential language-specific effects.

Neural attention tracking was analyzed using speech-to-EEG correlation-based methods \cite{alickovic2019tutorial, geirnaert_time-adaptive_2022, brodbeck_eelbrain_2023}, specifically temporal response functions (TRFs) - linear filters fitted between speech features and simultaneously recorded EEG signals. This allowed us to examine both qualitative TRF characteristics and the ability to decode participants’ attention across the three conditions and two EEG modalities. By combining scalp and cEEGrid recordings with naturalistic and dynamic listening tasks, we directly evaluated how AAD performs in realistic scenarios, addressing the ecological, dynamic, and practical challenges of everyday listening.

The paper proceeds with a detailed description of the methods in Section~\ref{sec:met}, covering both data recording and data analysis. Section~\ref{sec:res} presents the results, which are then discussed in Section~\ref{sec:dis}. Finally, conclusions and future directions are provided in Section~\ref{sec:con}.

\section{Methods \& Materials}
\label{sec:met}

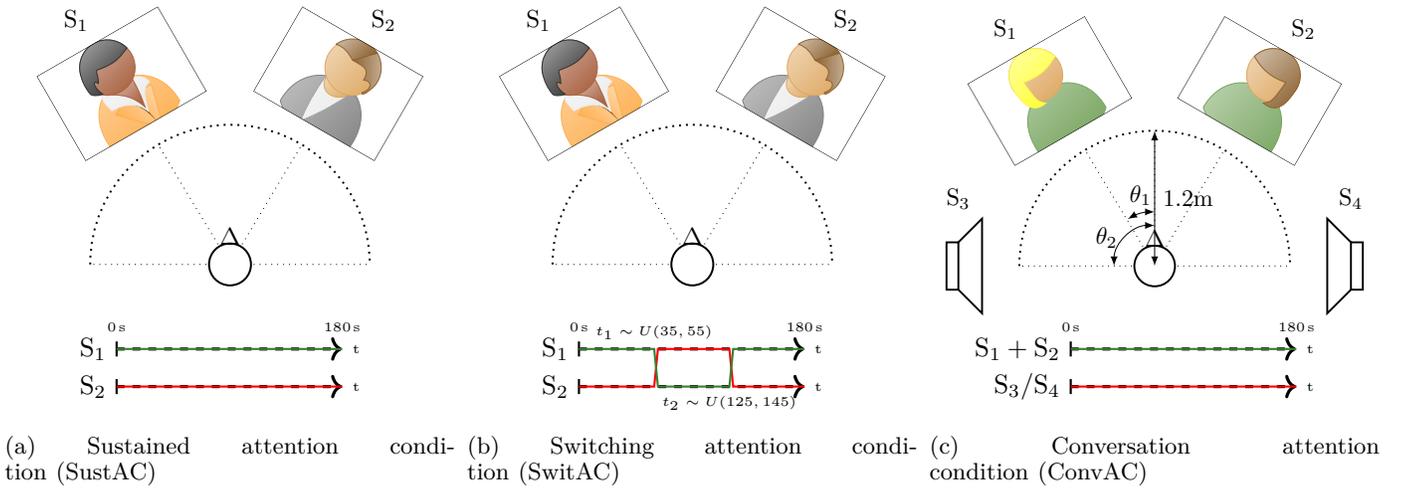
\begin{figure*}[t]
    \centering
    \begin{subfigure}[t]{0.33\textwidth}
        \centering
            \resizebox{\textwidth}{!}{
\begin{tikzpicture}
    \draw[thick,dotted] 
        (-2,0)  arc (180:0:2);
    \draw[dotted]
        (0,0) to (2,0)
        (0,0) to (-2,0)
        (0,0) to (1.029,1.715)
        (0,0) to (-1.029,1.715);
    \draw[thick,fill=white] 
        (0,0) circle(0.3);
\draw (0,0.4) node{$\Delta$};
\tvpersonleft{(-1.5,2.5)}{30}{alice};
\tvpersonright{(1.5,2.5)}{-30}{bob};
\speaker{(2.9,0)}{180}{0.7}{white};
\speaker{(-2.9,0)}{0}{0.7}{white};
\node at (-2.2,3.5) {$\text{S}_1$};
\node at (2.2,3.5) {$\text{S}_2$};
\end{tikzpicture}
} \\
            \vspace{-0em}
            \begin{tikzpicture}
    \pgfmathsetmacro{\h}{0.5}
    \pgfmathsetmacro{\hd}{0.4}
    \pgfmathsetmacro{\hu}{0.6}
    \draw[->,dashed,very thick] (0,0) node[left] {$\text{S}_2$} -- (3,0) node[right] {\tiny t};
    \draw[->,dashed,very thick] (0,\h) node[left] {$\text{S}_1$} -- (3,\h) node[right] {\tiny t};
    \draw[-,thick] (0,-0.1) -- (0,0.1);
    \draw[-,thick] (0,\hd) -- (0,\hu) node[above] {\tiny \SI{0}{\second}};
    \draw[-,thick,white] (3,\hd) -- (3,\hu) node[above,black] {\tiny \SI{180}{\second}};
    \draw[OliveGreen, thick]  (0,\h) -- (3,\h);
    \draw[red, thick]  (0,0) -- (3,0);
    \draw[white, thick, opacity=0]  (0,\h) -- (1,\h) node[above,black] {\tiny $t_1 \sim U(35,55)$} -- (1.05,0) -- (2,0) node[below,black] {\tiny $t_2 \sim U(125,145)$} -- (2.05,\h) -- (3,\h);
\end{tikzpicture}
            \label{fig:exp:a}
        \caption{\expandafter\MakeUppercase \sustname~(\sustnameshort)}
    \end{subfigure}%
    ~
    \begin{subfigure}[t]{0.33\textwidth}
        \centering
            \resizebox{\textwidth}{!}{
\begin{tikzpicture}
    \draw[thick,dotted] 
        (-2,0)  arc (180:0:2);
    \draw[dotted]
        (0,0) to (2,0)
        (0,0) to (-2,0)
        (0,0) to (1.029,1.715)
        (0,0) to (-1.029,1.715);
    \draw[thick,fill=white] 
        (0,0) circle(0.3);
\draw (0,0.4) node{$\Delta$};
\tvpersonleft{(-1.5,2.5)}{30}{alice};
\tvpersonright{(1.5,2.5)}{-30}{bob};
\speaker{(2.9,0)}{180}{0.7}{white};
\speaker{(-2.9,0)}{0}{0.7}{white};
\node at (-2.2,3.5) {$\text{S}_1$};
\node at (2.2,3.5) {$\text{S}_2$};
\end{tikzpicture}
} \\
            \vspace{-0em}
            \begin{tikzpicture}
    \pgfmathsetmacro{\h}{0.5}
    \pgfmathsetmacro{\hd}{0.4}
    \pgfmathsetmacro{\hu}{0.6}
    \draw[->,dashed,very thick] (0,0) node[left] {$\text{S}_2$} -- (3,0) node[right] {\tiny t};
    \draw[->,dashed,very thick] (0,\h) node[left] {$\text{S}_1$} -- (3,\h) node[right] {\tiny t};
    \draw[-,thick] (0,-0.1) -- (0,0.1);
    \draw[-,thick] (0,\hd) -- (0,\hu) node[above] {\tiny \SI{0}{\second}};
    \draw[-,thick,white] (3,\hd) -- (3,\hu) node[above,black] {\tiny \SI{180}{\second}};
    \draw[red, thick]  (0,0) -- (1,0) -- (1.05,\h) -- (2,\h) -- (2.05,0) -- (3,0);
    \draw[OliveGreen, thick]  (0,\h) -- (1,\h) node[above,black] {\tiny $t_1 \sim U(35,55)$} -- (1.05,0) -- (2,0) node[below,black] {\tiny $t_2 \sim U(125,145)$} -- (2.05,\h) -- (3,\h);
\end{tikzpicture}
            \label{fig:exp:b}
        \caption{\expandafter\MakeUppercase \switname~(\switnameshort)}
    \end{subfigure}%
    ~
    \begin{subfigure}[t]{0.33\textwidth}
        \centering
        \resizebox{\textwidth}{!}{
\begin{tikzpicture}
    \draw[thick,dotted]  
            (-2,0)  arc (180:0:2);
    \draw[dotted]
        (0,0) to (2,0)
        (0,0) to (-2,0)
        (0,0.5) to (0,2)
        (0,0) to (1.029,1.715)
        (0,0) to (-1.029,1.715);
    \draw[thick,fill=white] 
        (0,0) circle(0.3);
    \draw[latex-latex]  (90:0.8) arc(90:120:0.8) node[midway,above]{$\theta_1$};
    \draw[latex-latex]  (0,0.6) arc(90:180:0.6) node[midway,left]{$\theta_2$};
    \draw[latex-latex] (0,2) -- (0,0) node[midway,right]{$1.2\text{m}$};
    
    \draw (0,0.4) node{$\Delta$};
\tvpersonleft{(-1.5,2.5)}{30}{person,female,hair=yellow}
\tvpersonright{(1.5,2.5)}{-30}{person,female}
\node at (-2.2,3.5) {$\text{S}_{1}$};
\node at (2.2,3.5) {$\text{S}_{2}$};
\speaker{(2.9,0)}{180}{0.7}{black}
\speaker{(-2.9,0)}{0}{0.7}{black}
\node at (-2.9,1) {$\text{S}_3$};
\node at (2.9,1) {$\text{S}_4$};
\end{tikzpicture}
}\\
        \vspace{-0em}
        \begin{tikzpicture}
    \pgfmathsetmacro{\h}{0.5}
    \pgfmathsetmacro{\hd}{0.4}
    \pgfmathsetmacro{\hu}{0.6}
    \draw[->,dashed,very thick] (0,0)  node[left] {$\text{S}_3 / \text{S}_4$} -- (3,0) node[right] {\tiny t};
    \draw[->,dashed,very thick] (0,\h)  node[left] {$\text{S}_1+\text{S}_2$} -- (3,\h) node[right] {\tiny t};
    \draw[-,thick] (0,-0.1) -- (0,0.1);
    \draw[-,thick] (0,\hd) -- (0,\hu) node[above] {\tiny \SI{0}{\second}};
    \draw[-,thick,white] (3,\hd) -- (3,\hu) node[above,black] {\tiny \SI{180}{\second}};
    \draw[red, thick]  (0,0) -- (3,0);
    \draw[OliveGreen, thick]  (0,\h) -- (3,\h);
    \draw[white, thick, opacity=0]  (0,\h) -- (1,\h) node[above,black] {\tiny $t_1 \sim U(35,55)$} -- (1.05,0) -- (2,0) node[below,black] {\tiny $t_2 \sim U(125,145)$} -- (2.05,\h) -- (3,\h);
\end{tikzpicture}
        \label{fig:exp:c}
        \caption{\expandafter\MakeUppercase \convname~(\convnameshort)}
    \end{subfigure}%
    \caption{Experiment paradigm and condition designs. Top panels (a), (b), and (c) show the experimental setup of \sustnameshort, \switnameshort~and \convnameshort, respectively. The setups for \sustnameshort (a) and \switnameshort (b) are identical, with attention directed to one out of the two frontal AV speakers. In (c), attention is directed either to the two frontal AV speakers engaged in conversation or to the side single speaker. Below each setup, the instructed attentional focus over the course of a trial is shown: attention is sustained on a single talker for \sustnameshort and \convnameshort, whereas \switnameshort includes two switches between speakers.}
    \label{fig:exp}
\end{figure*}

The experimental protocol was approved by the ethics committee for the capital region of Denmark (journal number F-24047175), and by the Swedish Ethical Review Authority, Sweden (DNR: 2022-05129-01). The study was conducted according to the Declaration of Helsinki, and all the participants gave a written consent prior to the experiment.
\subsection{Study Population}
The study included 24 native Danish speakers (12 male), aged 23-51 (mean $\pm$ SD: $35.7 \pm 8.9$).
All participants had normal hearing, confirmed by otoscopy and audiogram measurements at 250, 500, 1000, 2000, 4000, and \SI{8000}{\hertz} for each ear, with hearing thresholds below \SI{25}{\decibel} HL. Fifteen participants wore glasses or contact lenses.
\subsection{Paradigm \& Experimental Design}
The experiment took place in a sound-proof room under controlled light conditions.
Participants sat in a chair in the middle of the room and were presented with AV stimuli from loudspeakers (Genelec 8010A) positioned at $\pm30^{\circ} (\theta_1)$ and $\pm 90^{\circ}(\theta_2)$ azimuth relative to the participants. The two speakers in front ($\text{S}_1$, $\text{S}_2$) had computer screens (Thinkvision, \SI{25}{''}), situated directly in front of the speakers. The distance from screens to the speakers were \SI{1.36}{\meter}. 
Loudspeakers were calibrated at \SI{65}{\decibel} SPL, using single speech with quiet parts longer than \SI{0.8}{\second} cut away. All stimuli was presented with a sample rate of \SI{44.1}{\kilo\hertz}. 
Loudspeakers to each side of the participant ($\text{S}_3$, $\text{S}_4$), used in the final experimental condition, had no screens as the participant were never instructed to look at these speakers.
A screen at $0^{\circ}$ azimuth, positioned at the same distance as the speaker screens, was used to present instructions to the participant.
Chair heights were adjusted so that each participant’s vertex aligned naturally with the screens.
A schematic of the setup is shown in Figure~\ref{fig:exp}. 

\subsubsection{Audio-visual stimuli}
Two sets of AV stimuli were used in the study, the single talker set and the conversation AV set. 
The single talker set consisted of excerpts from a Danish TV program with various known personalities sitting and answering questions posed to them from viewers through a live chat.
Throughout the trials, speakers primarily looked directly into the camera, with only brief deviations.
Four speakers (2 male and 2 female) made up the single talker AV set, discussing various topics related to their professional background.

The conversation AV set consisted of a podcast, with two female speaker engaged in conversation, looking at each other. 
The cameras were positioned to the side of each speaker, so that they faced one another in the frame. 
Discussion topics included names and friends. For the conversation set, audio from the non-target talker was attenuated in the the opposite talker's loudspeaker to achieve realistic spatialization. When both talkers spoke simultaneously, audio was presented through both speakers. Detailed information about both audiovisual sets is provided in the supplementary material.

\subsubsection{Procedure \& Conditions}

Participants completed three tasks across three blocks corresponding to the conditions: \sustname (\sustnameshort), \switname (\switnameshort), \convname (\convnameshort) (see Fig. \ref{fig:exp}). Each trial lasted \SI{180}{\second}.
\paragraph{Trial protocol} A training trial was completed for each condition to ensure participants understood the task. Each trial started with a task instruction, followed by stimuli. At the end of each trial, participants answered three two-choice questions related to the content of the attended stimuli and rated the listening and understanding difficulty on a 1–7 scale. Self-determined breaks were allowed after every second trial. Task instructions were presented visually (arrow on the center screen) and with a green dot below the attended speaker's face.   

\paragraph{\sustnameshort} Eight trials consisted of two competing single-talkers from the single talker set ($\text{S}_1$ and $\text{S}_2$) presented audiovisually  (Fig. \ref{fig:exp}a). Participants were instructed to maintain attention on one speaker for the entire trial while ignoring the other. A green attention target circle remained below the face of the attended speaker throughout the trial. Participants were asked to follow the attended speaker naturally with eye movement, as they would in real-life.

\paragraph{\switnameshort} Also consisting of 8 trials, the setup was similar to \sustnameshort, but participants switched attention between the two speakers ($\text{S}_1$ and $\text{S}_2$) twice per trial (Fig. \ref{fig:exp}b). Switch times were randomly drawn from the uniform distribution $\text{\textit{U}}(35,55)$ seconds and $\text{\textit{U}}(125,145)$ seconds, respectively. Switches were cued by changing the green attention target circle to yellow two seconds before the switch; at switch onset the circle moved to the other screen and reverted to green. Center-screen instructions were updated to indicate the newly attended speaker. Participants were instructed to naturally follow the currently attended speaker with their eyes and to shift gaze to the newly attended speaker after each switch.

\paragraph{\convnameshort} The setup differed, presenting a two-speaker conversation from the conversation set at $\text{S}_1$ and $\text{S}_2$, alongside a competing single speaker from the single talker set ($\text{S}_3$ or $\text{S}_4$), evenly randomized across trials (Fig. \ref{fig:exp}c). Participants were instructed to attend either the conversation or the side single talker. When attending the single talker, participants maintained fixation on a point on the center screen, whereas for conversation trials, they could naturally follow the speakers with eye movements while minimizing unnecessary head movements. \convnameshort consisted of 9 trials, with the extra trial always being attended towards the conversation. This was to increase the amount of data available to perform intra-conversation attention analysis.

For all conditions, competing speakers were always of opposite sex. Stimuli were randomized across participants to avoid biases. Within each condition, the distribution of attention toward male and female speakers, as well as left and right speaker positions, was balanced. After completing all three blocks, participants completed a questionnaire to report any familiarity with the speakers and rate the overall interestingness of each speaker. 
For SwitAC, the post-trial comprehension questions were based on speech segments from 0-\SI{30}{\second}, 60-\SI{120}{\second}, and 150-\SI{180}{\second} to avoid overlap with attention switches and cover material from both speakers. 

\begin{figure}
    \centering
    \input{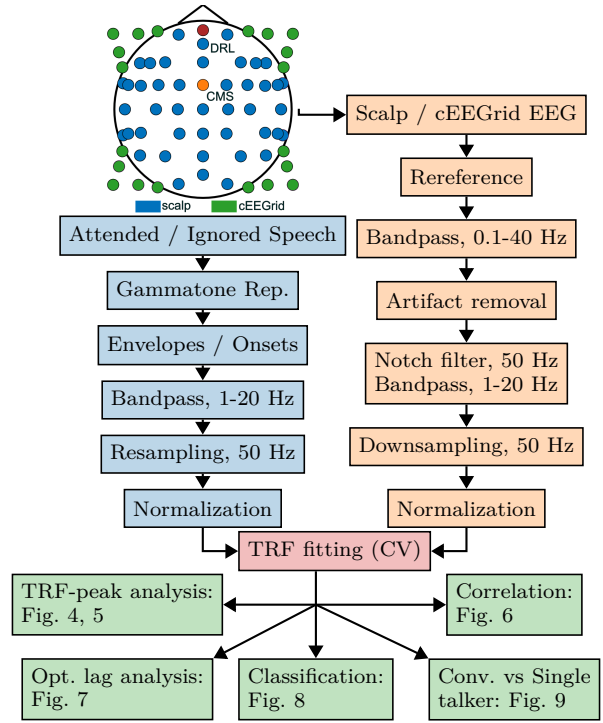}
    \caption{The EEG setup with common mode sense (CMS) and driven right leg (DRL) electrodes, and subsequent preprocessing of stimuli (blue) and EEG data (orange) are shown as flowcharts. Forward and backward temporal response functions (TRF) models are then fitted using cross-validation (CV) (red). The forward model is analyzed both qualitatively and quantitatively by inspecting TRF waveforms, while the backward model is analyzed quantitatively via correlation-based metrics, including reconstruction accuracy and classification performance (green).}
    \label{fig:dataflow}
\end{figure}
\subsection{Data Acquisition}
\subsubsection{Neural Data}

Neural responses were recorded through  64-channel EEG, with a modified Easycap and a Smarting Pro X amplifier. To include cEEGrid, 20 cap electrodes were replaced with a pair of 10-channel around-the-ear arrays. This resulted in 44 scalp EEG channels (blue in Figure~\ref{fig:dataflow}) and 20 cEEGrid channels ({\footnotesize L1-10} and {\footnotesize R1-10}, green in Figure~\ref{fig:dataflow}). 

EEG and other measurements were digitally recorded through the associated mbtStreamer application.  The sampling rate of EEG was \SI{500}{\hertz}, with {\footnotesize FCZ} as reference electrode. Impedances were visually monitored to remain be below \SI{20}{\kilo\ohm}.

\subsubsection{Other data streams}
Two microphones (Behringer ECM-8000) were placed \SI{10}{\centi\meter} behind each of the participant's ears to capture the sound field  (not analyzed in this study). Sound was preamplified (Golden Age Project PRE-73 Jr MKII) before digital conversion using Ferrofish PULSE16 MX, and subsequent downsampled to \SI{11025}{\hertz}. Audio was simultaneously captured and sent to the mbtStreamer application. A Tobii Pro Nano eye-tracking device \cite{tobii} was positioned in front of the participant; this data is also out of the scope of this study. 

\subsubsection{Data-stream synchronization}
The experiment was run on Windows using the Psychopy package for Python \cite{peirce_psychopy2_2019}. Sound I/O was handled through the Ferrofish PULSE16 MX.
Stimuli and data synchronization were performed on a separate computer using LabStreamingLayer \cite{noauthor_sccnlabstreaminglayer_2025}, which aligned triggers with EEG and other measured time series across devices over the local network.

\subsection{Data Preprocessing}
Neural data was initially separated into scalp EEG and cEEGrid EEG. Scalp EEG was rereferenced to average of all channels, and cEEGrid EEG was rereferenced to the {\footnotesize L4} and {\footnotesize R4} channels \cite{mirkovic_target_2016}. All subsequent preprocessing steps were performed in parallel for both EEG sets (see flowcharts in Fig. \ref{fig:dataflow}). Preprocessing included notch-filtering at \SI{50}{\hertz} to remove line noise, followed by bandpass-filtering between 0.1-\SI{40}{\hertz} to remove EEG drifts and unwanted noise. Independent component analysis (ICA) using with the extended infomax algorithm \cite{lee1999independent} was then applied with 20 components for scalp EEG and 9 components for cEEGrid EEG. Manual inspection and removal of artifactual components resulted in an average of 5 components removed for scalp EEG and 3 for cEEGrid EEG. After component reprojection, EEG data were bandpass-filtered between 1-\SI{20}{\hertz}, downsampled to \SI{50}{\hertz}, and normalized (see Fig. \ref{fig:dataflow}).

Speech feature preprocessing began with Gammatone (GT) filtering of the presented audio. At each timepoint, the GT representations included 128 frequency bands spaced according to the equivalent rectangular bandwidth scale between \SI{80}-\SI{15000}{\hertz}, with \SI{1}{\milli\second} temporal resolution, implemented in Eelbrain \cite{brodbeck_eelbrain_2023}. Two representations were derived: the \textbf{acoustic envelope}, computed as the sum of the absolute values of the GT representation across frequency bands, and \textbf{acoustic onsets}, extracted using the acoustic edge detection method \cite{fishbach_auditory_2001, brodbeck_neural_2020}, implemented in \cite{brodbeck_eelbrain_2023}. This approach was compared to an alternative method based on the half-wave rectified derivative with a Savitzky–Golay filter, which yielded similar results and the half-wave rectified method was therefore omitted. Both speech representations were subsequently bandpass-filtered between 1-\SI{20}{\hertz}, downsampled to \SI{50}{\hertz}, and normalized.
\subsection{Data Analysis}

Neural tracking of speech was modeled using the \textbf{temporal response function} (TRF), a linear finite impulse response model \cite{o2015attentional, alickovic2019tutorial}. Both forward and backward TRF models were used to relate time-varying speech features to concurrent EEG signals. 
Let the TRF for channel $i$ be $\mathbf{h}(i)  = [h(l_1,i),h(l+1,i),...,h(l_2,i)]^T$ over time-lags $l_1$ to $l_2$ chosen between cut-off time-lags $t_{min}, t_{max}$. 
The forward TRF model  predicts the EEG signal  
$\hat{y}_i(k)$ for channel $i$ from the speech feature $x(k)$:
\begin{equation}
    \label{eq:forward}
    \hat{y}_i(k) = \sum_{l=l_1}^{l_2} h(l,i) x(k-l).
\end{equation}
The backward TRF model reconstructs the speech feature  
$\hat{x}(k)$ from the EEG signal $y_i(k)$:
\begin{equation}
    \label{eq:backward}
    \hat{x}(k) = \sum_{i=1}^{n_{ch}} \sum_{l=l_1}^{l_2} h(l,i) y_i(k+l).
\end{equation}
TRF estimation in Eq. \eqref{eq:forward} and \eqref{eq:backward} was performed using the boosting algorithm \cite{zhang_boosting_2005, david_estimating_2007} in Eelbrain \cite{brodbeck_eelbrain_2023}, which promotes sparse, interpretable solutions. Robustness was further enhanced by expressing the TRF as a weighted sum of basis functions, $h(l,i) = \sum_{p=1}^P w_p \; \phi_p(l)$, where basis weights $w_p$ are estimated instead of raw TRF coefficients $h(l,i)$. The basis functions $\phi_p$ consist of Hamming windows with \SI{50}{\milli\second} width, centered around each element of the TRF. Filters were fitted using time-lags from -\SI{1}{\second} to \SI{1}{\second}, minimizing the mean absolute error of each regression problem

\subsubsection{Correlation Metrics and Optimal Lag Analysis}
 
Correlations between model estimates and measured signals are consistently stronger for attended speech than for ignored speech, providing a robust marker of selective attention \cite{o2015attentional}. In the forward model (Eq. \eqref{eq:forward}), predictions of EEG activity were correlated with recorded EEG, whereas in the backward model (Eq. \eqref{eq:backward}), reconstructions of the speech envelope were correlated with the actual speech stimuli. Correlations were computed using the Pearson correlation coefficient:

\begin{equation}
    \rho_{\mathbf{x}, \mathbf{y}}=
    \frac{\sum_k (x(k)-\bar{x}) (y(k)-\bar{y})}{\sqrt{\sum_k (x(k)-\bar{x})^2 \sum_k (y(k)-\bar{y})^2}}
\end{equation}

Optimal time lag analysis found the lag-intervals with the highest reconstruction/prediction correlations. Replicating the method in \cite{mirkovic_target_2016}, forward and backward filters were fitted for multiple sliding time lag windows of length \SI{45}{\milli\second}, with \SI{30}{\milli\second} overlap, in a range from \SI{-600}{\milli\second} to \SI{600}{\milli\second}. Correlation and performance metrics were averaged across cross-validation (CV) folds and EEG channels (for the forward model), resulting in one estimate per subject for each sliding window.

\subsubsection{Statistical Testing and Cross-Validation}
Tracking auditory attention can further be improved by analyzing components of the TRF waveforms. TRFs estimated from attended speech exhibit distinct components compared to those estimated from ignored speech, reflecting underlying cognitive processes engaged during attention. These components can be statistically compared to identify significant differences between predictions from attended versus ignored speech. Mass-univariate statistics were employed using independent samples t-tests \cite{brodbeck_eelbrain_2023}, with threshold-free cluster enhancement correcting for multiple comparisons \cite{Smith2009-fw}. TRFs were smoothed with Gaussian kernels of width \SI{50}{m\second} to improve sensitivity. To test correlation metrics, paired sample t-tests were used, with multiple comparisons corrections employed to control for the family-wise error rate. 

All CV-based performance metrics were obtained using a leave-one-trial-out strategy to prevent overfitting. This approach avoids bias arising from temporal similarity within the same trial, which has been shown to affect auditory attention decoding performance \cite{tanveer2024deep, puffay2023relating}.

\section{Results}
\label{sec:res}

\subsection{Behavioral data analysis}

\begin{figure}[t]
     \centering
     \includegraphics[width=1\columnwidth]{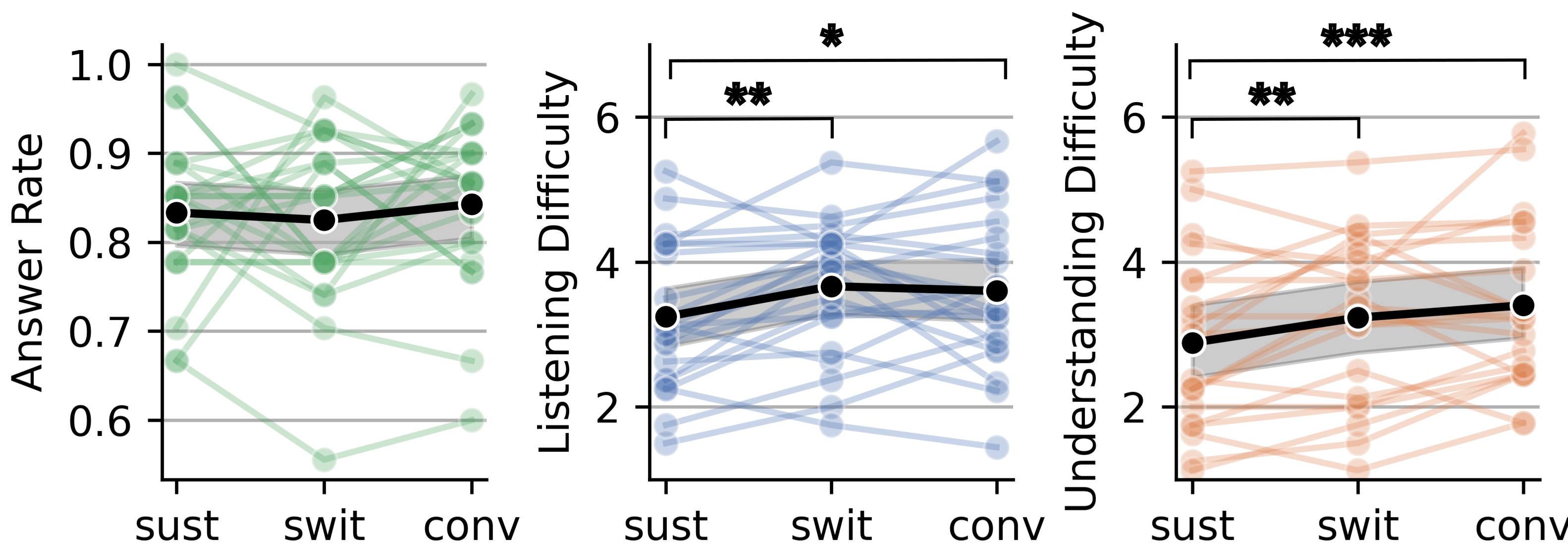}
    \caption{\textbf{Behavioral analysis:} Answer rate (left), self-rated listening difficulty (middle), and self-rated understanding difficulty (right) for the three conditions, SustAC, SwitAC and ConvAC. Colored dots show individual subject averages; black markers indicate the group mean. Ratings are on a $1-7$ scale. Significance was corrected for multiple comparisons using the Benjamini-Hochberg procedure \cite{benjaminihochberg}.}
    \label{fig:behav}
\end{figure}

To confirm that participants performed the listening tasks as instructed across the three conditions, the percentage of correctly answered questions was calculated (Figure~\ref{fig:behav}, left). Response accuracy was 85\% for \sustnameshort, 84\% for \switnameshort, and 89\% for \convnameshort. This indicates that participants focused on the attended talker and understood the speech material. 

Self-rated difficulty was recorded on a $1-7$ scale after each trial, separately for listening difficulty (Figure~\ref{fig:behav}, middle) and understanding difficulty (Figure~\ref{fig:behav}, right), with 1 representing the lowest difficulty. Individual averages are shown in color, and overall averages in black. Paired t-tests revealed a significant increase in listening difficulty from \sustnameshort to \switnameshort and \convnameshort ($p <0.05$, $p<0.1$). Similarly, both \switnameshort and \convnameshort were rated as significantly more difficult to understand than \sustnameshort ($p <0.05$, $p <0.01$).

\subsection{Neural data analysis}

\begin{figure*}[t]
    \centering
    \includegraphics[width=0.95\linewidth]{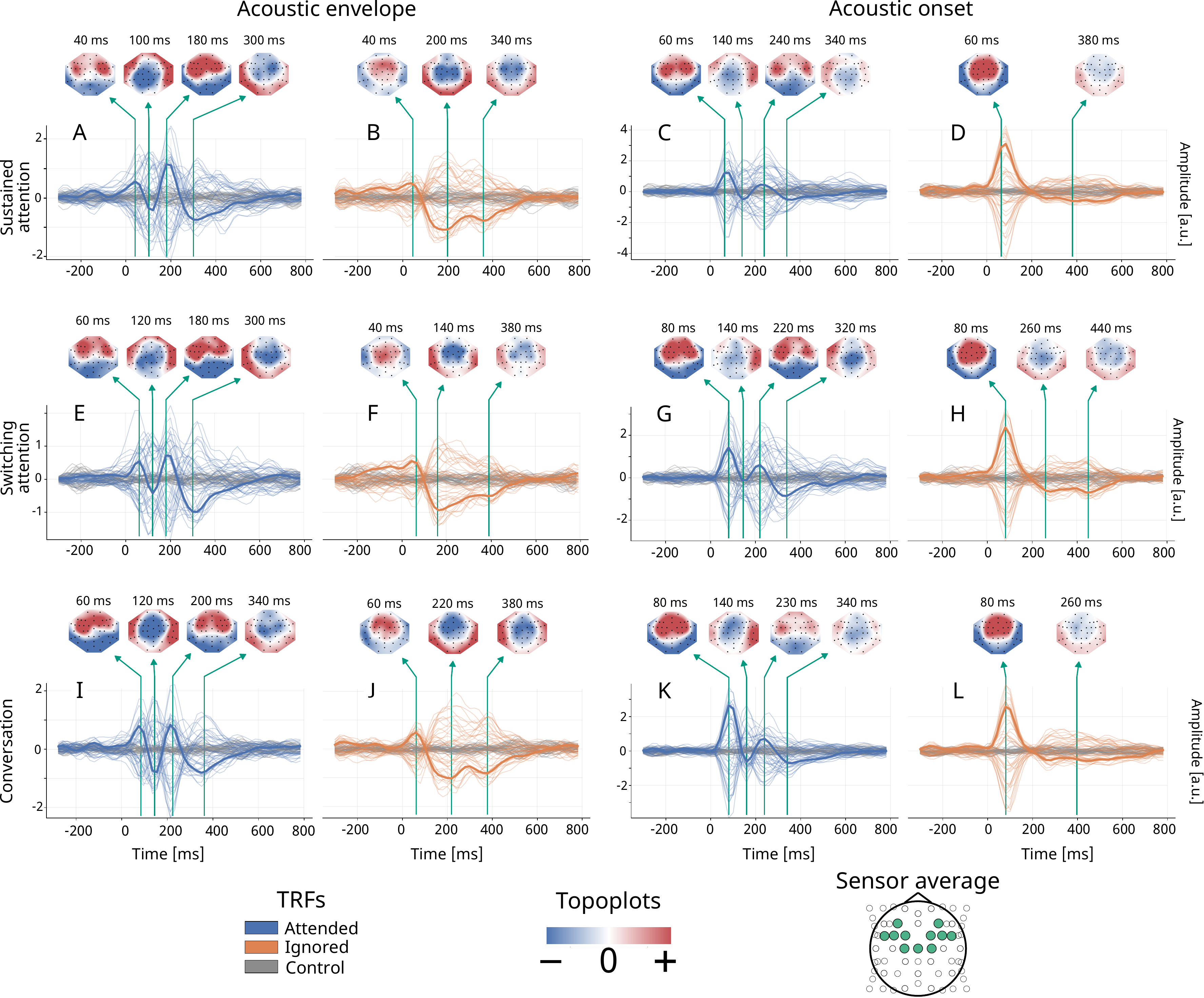}
    \caption{\textbf{TRF peak analysis.} TRFs for our three conditions \sustnameshort(A-D), \switnameshort(E-H), and \convnameshort(I-L), shown for attended (blue) and ignored (orange), with random control speech (grey). The left two and right two columns show TRFs for the acoustic envelope and the acoustic onset, respectively. The topographic plots show spatial patterns at observed peaks of interest, where red indicates positive amplitude and blue negative amplitude. All presented plots are for the scalp EEG sensors, with the thicker lines showing the TRF channel-average, as indicated by the cap layout at the bottom.}
    \label{fig:trf}
\end{figure*}

\begin{figure}[t]
    \centering
    \includegraphics[width=0.95\linewidth]{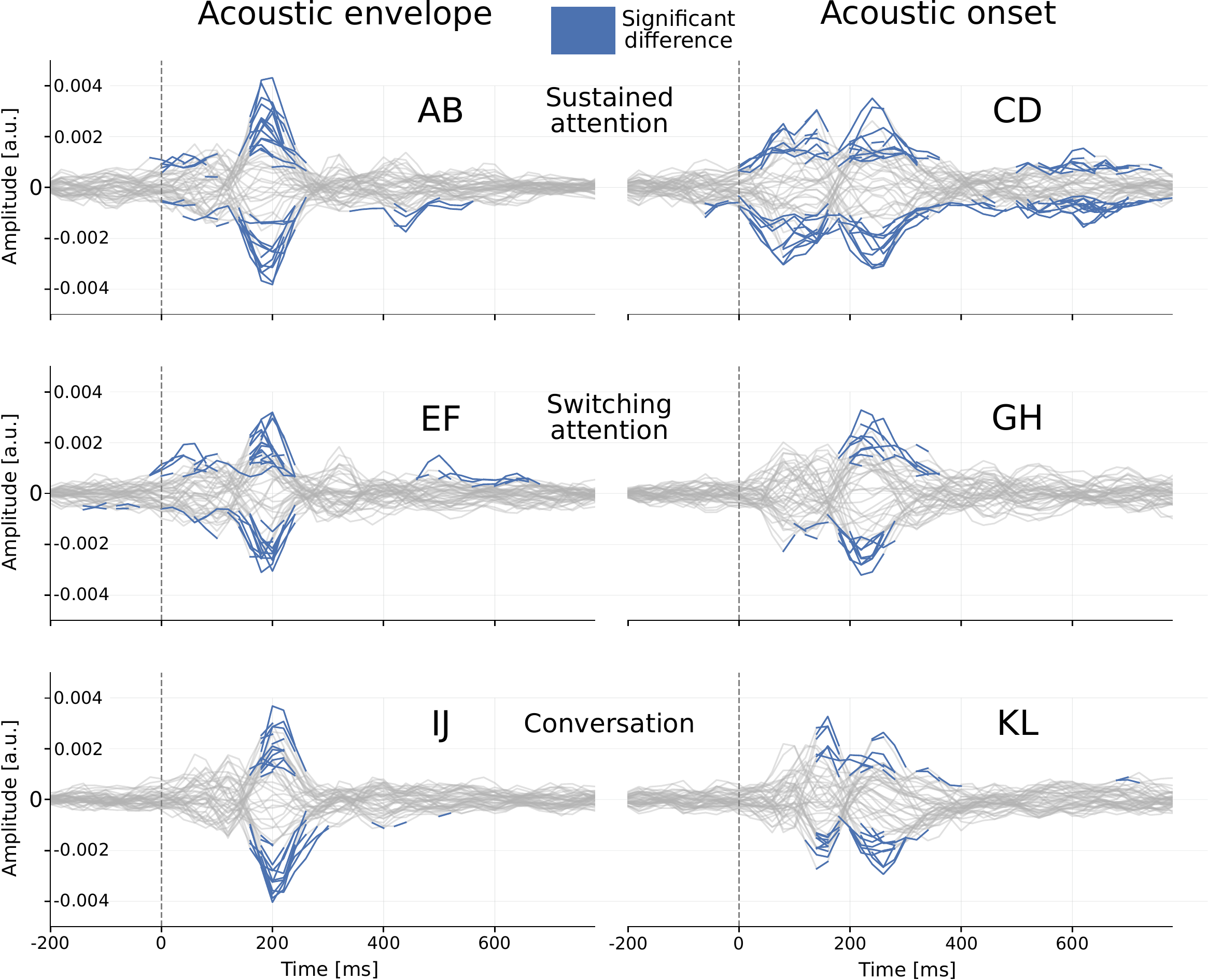}
    \caption{\textbf{TRF cluster analysis:} statistical cluster analysis on the differences between attended and ignored TRFs presented in Figure~\ref{fig:trf}. AB is the result from Figures~\ref{fig:trf}A and \ref{fig:trf}B, CD from \ref{fig:trf}C and \ref{fig:trf}D, etc. The analysis used an independent samples t-test in Eelbrain, where $p<0.05$ clusters are shown in blue.}
    \label{fig:stattrf}
\end{figure}

Neural responses to attended and ignored speech were analyzed using forward and backward models, and classification metrics. Figure~\ref{fig:trf} shows average forward TRFs across the three conditions (\sustnameshort, \switnameshort, \convnameshort) for attended (blue), ignored (orange), and control (gray) speech, with separate columns for the acoustic envelope and acoustic onset features. Topoplots illustrate scalp distributions at peaks of interest. Figure~\ref{fig:stattrf} presents statistical cluster analysis, highlighting significant differences between attended and ignored TRFs. 

Reconstruction correlations from backward models (based on \SI{35}{\second} segments) are presented in Figure~\ref{fig:recon}, and optimal lag analysis for forward and backward models  is shown in Figure~\ref{fig:optlag}. Attended vs. ignored speech classification results are presented in Figure~\ref{fig:class}. Lastly, trials in \convnameshort were separated between attention to a front conversation and attention to a side talker, with the resulting TRFs compared in Figure~\ref{fig:conv}. 

The results from the three conditions are discussed below.

\subsubsection{Sustained attention (\sustnameshort)}


Top row of Figure~\ref{fig:trf} shows the TRFs for \sustnameshort. Individual electrodes are shown as thin lines, with the average over 11 central-frontal sensors bolded.  For acoustic envelope (Figure~\ref{fig:trf}A,B), attended speech (Figure~\ref{fig:trf}A) shows peaks for P1$\sim$\SI{40}{\milli\second}, N1$\sim$\SI{100}{\milli\second}, and P2$\sim$\SI{180}{\milli\second}, with P2 being the largest. Ignored speech  (Figure~\ref{fig:trf}B) shows smaller amplitudes at these latencies, and cluster reveals significant differences at P2 and around 400 ms (Figure~\ref{fig:stattrf}AB, $p<0.05$), consistent with the N400 associated with higher-order linguistic processing and working memory capacity \cite{workingmemory1, workingmemory2, GillisENEURO.0075-23.2023}. 
For the acoustic onset (Figure~\ref{fig:trf}C,D), attended speech (Figure~\ref{fig:trf}C) shows similar peaks but shifted later: P1$\sim$\SI{60}{\milli\second}, N1$\sim$\SI{140}{\milli\second}, and P2$\sim$\SI{240}{\milli\second}, with P1 being the largest.
Ignored speech (Figure~\ref{fig:trf}D) shows smaller peak amplitudes, and cluster analysis identifies significant differences in peaks around \SI{100}{\milli\second}, \SI{240}{\milli\second}, and also at \SI{600}{\milli\second} (Figure~\ref{fig:stattrf}CD). The differences between the two speech features are  discussed in Section~\ref{sec:disc:features}.


Reconstruction accuracy (i.e., correlations between reconstructed and actual speech) from the backward model (Figure~\ref{fig:recon}A) showed significantly higher correlation for attended speech (blue) than for ignored speech (orange)  for scalp electrodes ($p<0.01$) with both speech features, while for cEEGrid electrodes, significance appeared only for acoustic onsets ($p<0.1$). Control speech (grey) was not tested.

Optimal lag analysis (Figure~\ref{fig:optlag}) was conducted to investigate temporal response characteristics,  evaluating the reconstruction/prediction accuracy of short $\SI{45}{\milli\second}$ TRF models across multiple time lags \cite{mirkovic_target_2016}.  As shown in Figure~\ref{fig:optlag}A for backward  (left) and forward (right) models, attended (blue) and ignored (orange) speech, scalp (solid lines) and cEEGrid (dashed lines) electrodes, the highest correlation occurred for attended speech with scalp electrodes at lag range $\tau \approx 40-\SI{200}{\milli\second}$ for both models, corresponding to P1, N1, and P2 peaks (Figure~\ref{fig:trf}A). A small, later rise around \SI{350}{\milli\second} reflected the N400 component (Figure~\ref{fig:stattrf}AB). The ignored speech with scalp electrodes (solid orange line) peaked marginally later (\SI{220}{\milli\second}). For cEEGrid electrodes, separation between attended and ignored speech started after \SI{100}{\milli\second} and peaked around \SI{250}{\milli\second}, differing from \cite{mirkovic_target_2016} and from our results for \switnameshort and \convnameshort.


Attention classification analysis (Figure~\ref{fig:class}) determined the successful decoding of attended versus ignored speech. Using the full trial duration ($\sim\SI{3}{\minute}$) resulted in the highest accuracy ($>80\%$) for scalp electrodes (solid lines). Performance for scalp sensors was robust, remaining above chance level even with $\SI{1.1}{\second}$ windows. The low cEEGrid performance (dashed lines) is discussed further in Section \ref{sec:disc:ceegrid}.

\subsubsection{Switching attention (\switnameshort)}


The middle row of Figure~\ref{fig:trf} shows the TRF-peak analysis for \switnameshort with acoustic envelope (E, F) and acoustic onset (G, H). Individual electrodes are thin lines, with the average over 11 central-frontal electrodes bolded.  The TRFs generally resemble those of \sustnameshort (Figures~\ref{fig:trf}A–D).  However, a consistent $\SI{20}{\milli\second}$ latency shift was observed, delaying both the P1 and N1 peaks for the acoustic envelope feature, and also delaying the P1 peak for the acoustic onset feature. This latency shift is discussed further in Section \ref{sec:dis}. Cluster analysis (Figure~\ref{fig:stattrf}E,F) reveals significant differences between attended (Figure~\ref{fig:trf}E, G) and ignored (Figure~\ref{fig:trf}F, H) speech, with the largest cluster around P2 ($\sim$\SI{190}{\milli\second}) and additional clusters covering P1 and N1. A higher-order latency cluster with opposite polarity appears between $\tau\sim 450-\SI{700}{\milli\second}$. For acoustic onset (Figure~\ref{fig:stattrf}G,H), the P2-peak latency is larger than for acoustic envelope, consistent with the corresponding TRFs in Figure~\ref{fig:trf}.

\begin{figure}[t]
    \centering
    \includegraphics[width=0.8\linewidth]{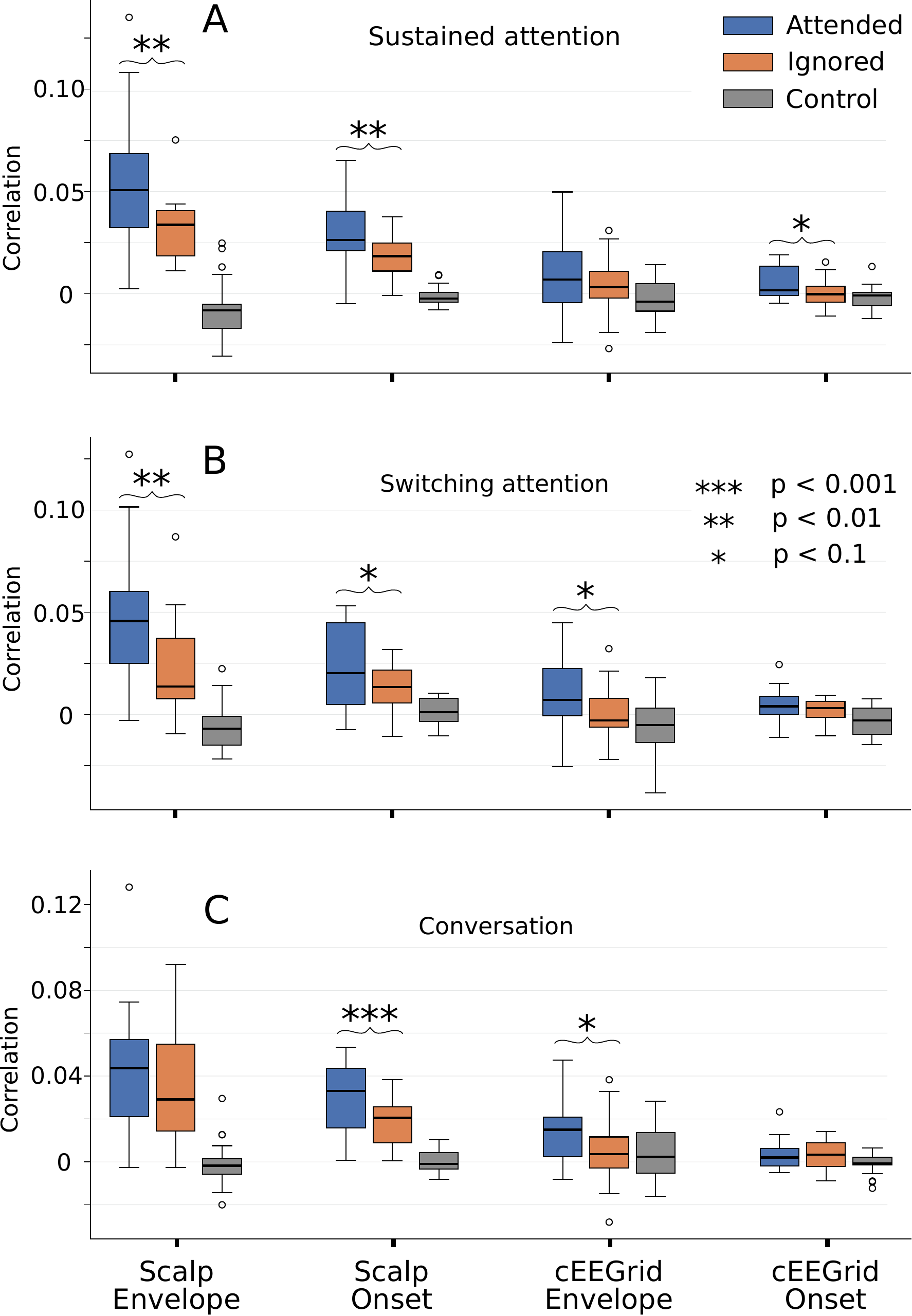}
    \caption{\textbf{Correlation Analysis:} Pearson’s correlation between reconstructed and real speech (backward model, \SI{35}{\second}) for \sustnameshort (A), \switnameshort (B), and \convnameshort (C). The $x$-ticks indicate electrode type (scalp, cEEGrid) and speech features (acoustic envelope, acoustic onset). Boxplots show attended (blue), ignored (orange), and control (grey) speech, with the median as the black line. Significance stars mark differences of means between attended and ignored speech. No statistical tests were performed between conditions and control speech.}
    \label{fig:recon}
\end{figure}

Reconstruction accuracy (Figure~\ref{fig:recon}B) confirmed the attentional effect  for \switnameshort. As in \sustnameshort (Figure~\ref{fig:recon}A), the attended speech (blue) had higher correlation than ignored (orange) speech and the control (grey) speech across all electrode types and speech features. Significant differences between attended and ignored speech were found for scalp electrodes with acoustic envelope ($p<0.01$) and acoustic onset ($p<0.1$), as well as for the cEEGrid electrodes with acoustic envelope ($p<0.1$). 

\begin{figure}[t]
    \centering
    \includegraphics[width=0.95\linewidth]{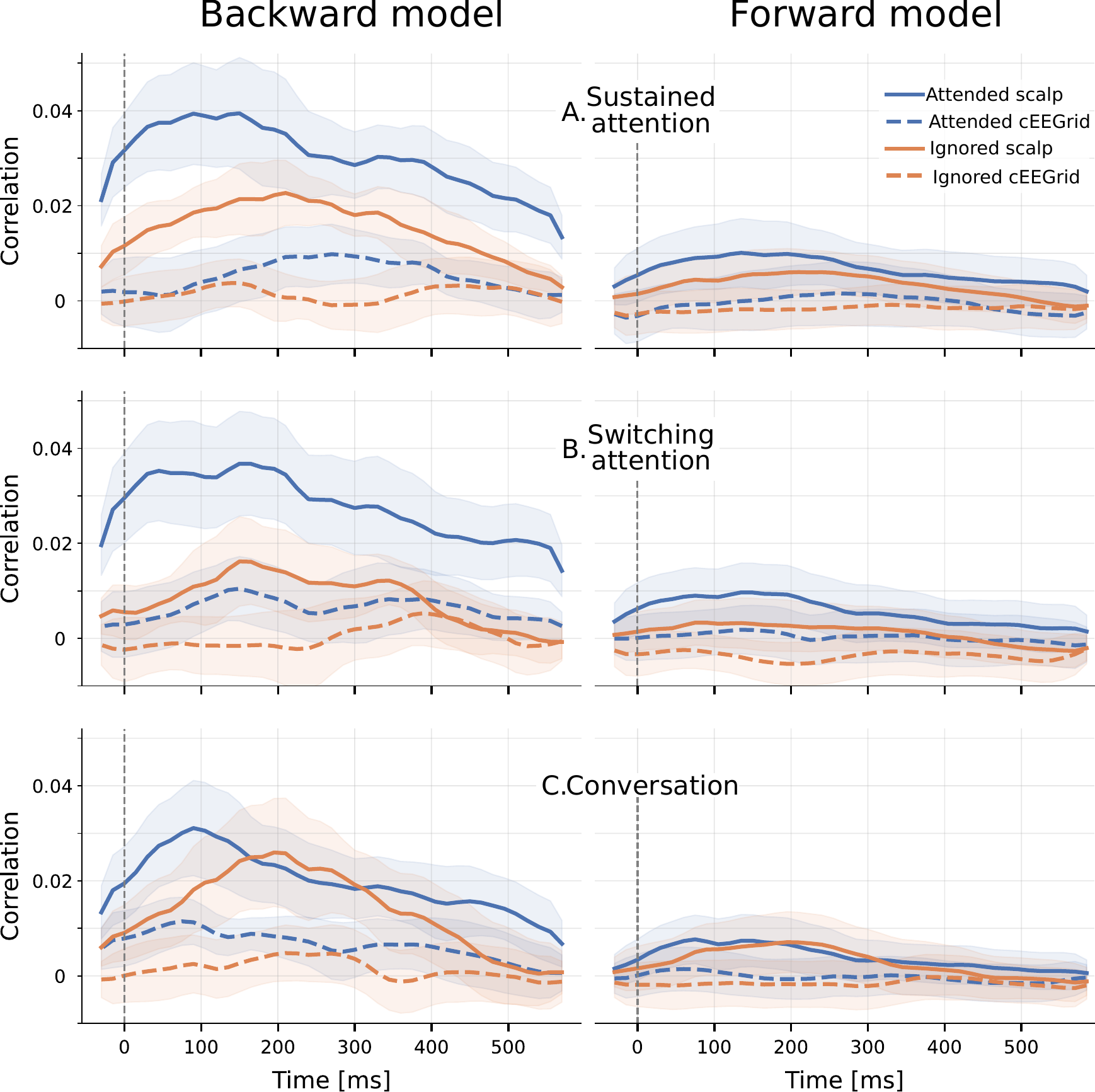}
    \caption{\textbf{Optimal lag analysis:} Pearson's correlation from backward (left) and forward (right) models for \sustnameshort (A), \switnameshort (B), and \convnameshort (C), using a \SI{45}{\milli\second} overlapping time-lag windows. Each plot shows the attended (blue) and ignored (orange) speech envelope, with scalp electrodes (solid lines) and cEEGrid electrodes (dashed lines). The $95\%$ confidence interval is shaded for each condition. Forward-model correlations are averaged across all 44 scalp and 20 cEEGrid electrodes.}
    \label{fig:optlag}
\end{figure}

The optimal lag analysis (Figure~\ref{fig:optlag}B) for scalp EEG with attended speech showed patterns similar to \sustnameshort. However, ignored speech and attended speech with cEEGrid, most clearly visible for the backward model, showed two distinct peaks ($\sim$\SI{150}{\milli\second} and $\sim$\SI{350}{\milli\second}), in contrast to the single dominant peak 200-\SI{300}{\milli\second} observed in \sustnameshort.

\begin{figure}[t]
    \centering
    \includegraphics[width=0.95\linewidth]{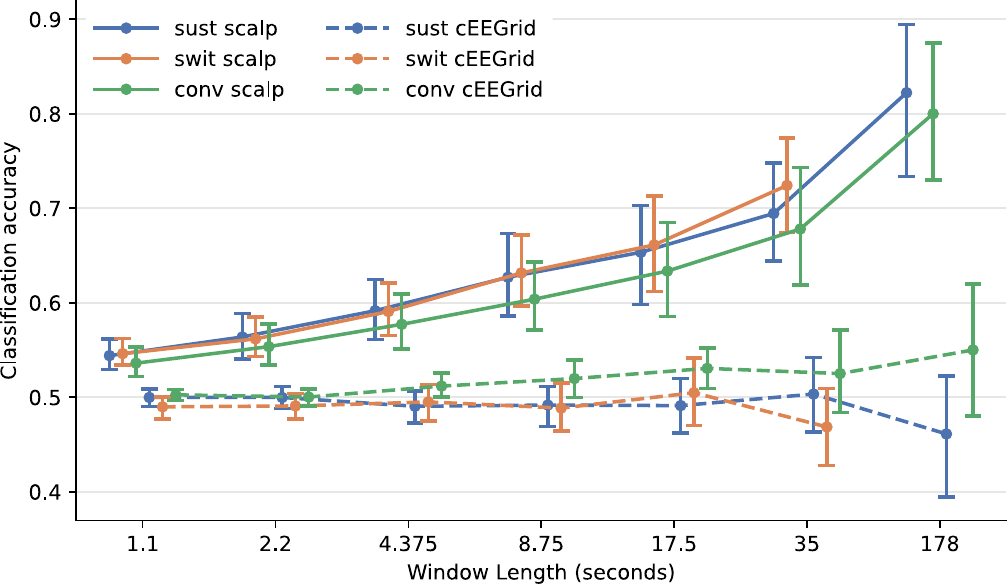}
    \caption{\textbf{Attention classification analysis:}  Average Pearson's correlations and confidence intervals across participants, for backward models applied on data of different lengths. For \switnameshort, no values are shown for the \SI{178}{\second} window, as attention is not sustained for that duration.}
    \label{fig:class}
\end{figure}

Attention classification analysis for the \switnameshort (Figure~\ref{fig:class}), orange) showed accuracy similar to \sustnameshort (blue). Accuracy for scalp electrodes (solid line) decreased from approximately $72\%$ at a $\SI{35}{\second}$ window to $55\%$ at a $\SI{1.1}{\second}$ window. Note that segment lengths were restricted and do not extend to $\SI{178}{\second}$, as attention was not sustained toward one talker for such a duration. The lower performance for cEEGrid electrodes (dashed line) will be further discussed in Section \ref{sec:disc:ceegrid}.

\subsubsection{Conversation attention (\convnameshort)}


The bottom row of Figure~\ref{fig:trf} shows the TRF-peak analysis for \convnameshort. Individual electrodes are thin lines, with the average over 11 central-frontal electrodes bolded.  For attended speech with acoustic envelope (Figure~\ref{fig:trf}I), four peaks showed similar amplitudes, unlike P2-dominant pattern in \sustnameshort (Figure~\ref{fig:trf}A). For the acoustic onset (Figure~\ref{fig:trf}K), a strong P1 dominated.
Peak latencies generally resembled \switnameshort, with P1 and N1 delayed by $\sim$\SI{20}{\milli\second} relative to \sustnameshort. Cluster analysis confirmed significant attended-ignored differences: a large cluster centered around P2 ($\sim$\SI{200}{\milli\second}) for the envelope (Figure~\ref{fig:stattrf}IJ), and clusters at $\sim$\SI{160}{\milli\second} and $\sim$\SI{250}{\milli\second} for the onset (Figure~\ref{fig:stattrf}KL).

\begin{figure}[t]
    \centering
    \includegraphics[width=0.9\linewidth]{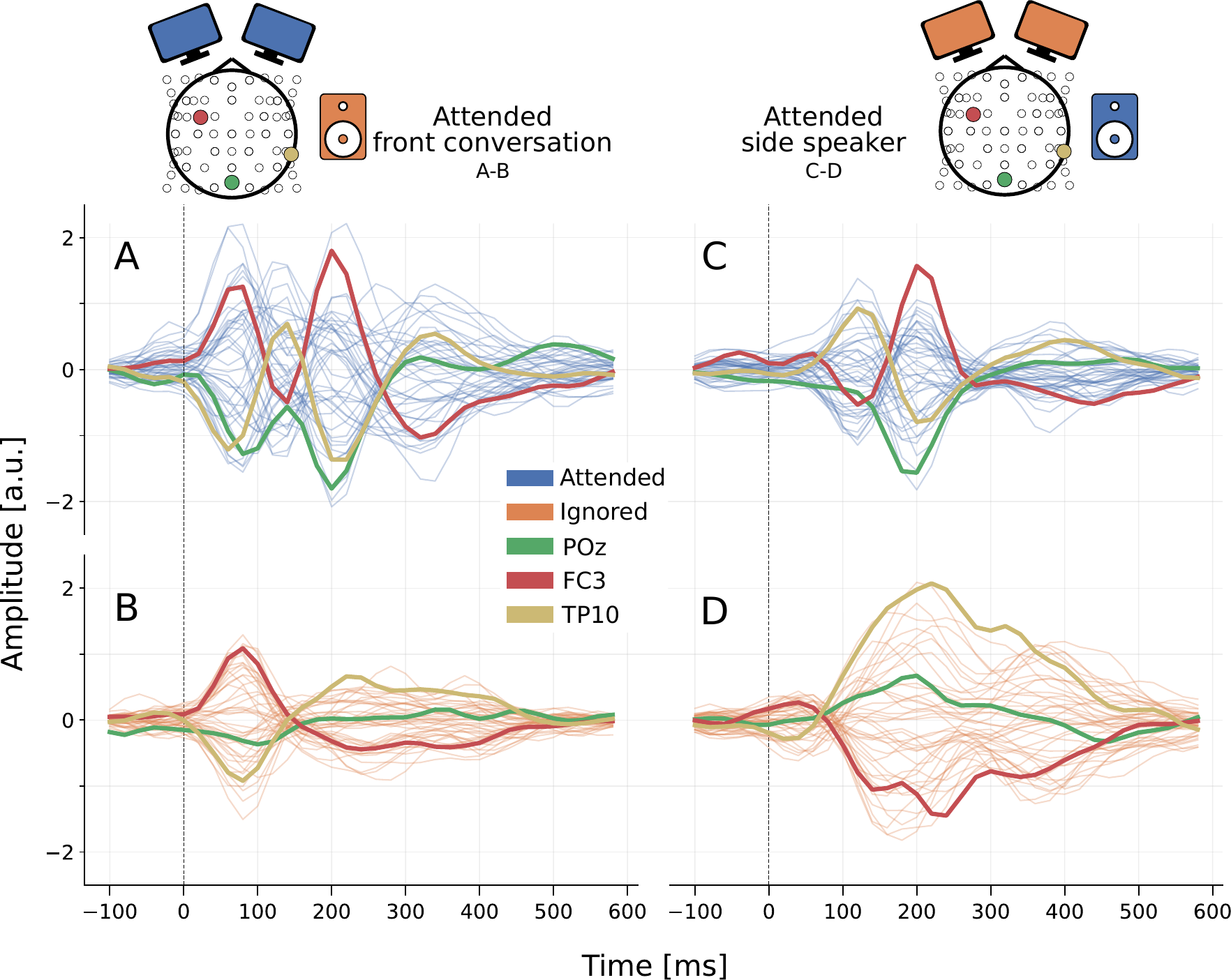}
    \caption{\textbf{Attention to conversation vs. side talker:} \convnamemid~analysis for trials directed to the front conversation (A-B) and to the side speaker (C-D). Attended speech is shown in blue (A, C) and ignored speech in orange (B, D). Three sensors of interest are highlighted: {\footnotesize POZ, FC3} and {\footnotesize TP10}. All TRFs use the acoustic envelope using scalp EEG.}
    \label{fig:conv}
\end{figure}

Figure~\ref{fig:conv} compares TRFs for trials separating attention to the front conversation (left) versus the side talker (right).  The top and bottom row shows TRFs for attended  and ignored speech respectively, all for the acoustic envelope speech feature and the scalp EEG sensors. Three interesting sensors are highlighted for deeper analysis in Section~\ref{sec:dis}: {\footnotesize POZ, FC3} and {\footnotesize TP10}. There is a clear similarity of the TRFs between A and D, as for B and C. This indicates that the TRFs trained on conversation features share commonalities compared to single speech, regardless of attentional state, which is further discussed in Section \ref{sec:disc:singleVSconv}.

Figure~\ref{fig:recon}C shows correlations between reconstructed and real speech (i.e., reconstruction accuracy) for \convnameshort, comparing attended (blue), ignored (orange), and control (grey) speech. Median correlations are indicated by black lines, shown for scalp and cEEGrid electrodes and two speech features (acoustic envelope, acoustic onset). Attended speech had significantly higher mean correlations than ignored speech for scalp EEG with acoustic onset ($p<0.001$) and for cEEGrid with acoustic envelope ($p<0.1$). Interestingly, for scalp EEG with acoustic envelope, the median correlation for attended speech was similar to \switnameshort ($\approx$0.043), but the ignored speech median was higher, although showing no significant difference in mean.

Figure~\ref{fig:optlag}C presents the optimal lag analysis. For scalp sensors in the backward model, attended speech (blue, solid) peaked around \SI{100}{\milli\second}, while  ignored speech (orange, solid) peaked later around \SI{200}{\milli\second} in both backward and forward models, consistent with attention-related P2 response.
Notably, the ignored correlations exceeded attended correlations between 180–\SI{300}{\milli\second}.
For cEEGrid electrodes, correlation of attended speech (dashed blue line) was consistently above ignored speech for the backward model, with peaks at \SI{90}{\milli\second}, \SI{180}{\milli\second}, and \SI{350}{\milli\second}. The largest peak for cEEGrid (correlation $\approx$0.01) occurs earlier for \sustnameshort$\sim$\SI{280}{\milli\second}, compared to \switnameshort$\sim$\SI{150}{\milli\second}, which also occurs earlier than \convnameshort$\sim$\SI{90}{\milli\second}. A small separation between attended (dashed blue line) and ignored (dashed orange line) speech is visible in the forward model, though correlations remain low.


The classification analysis for the \convnameshort in Figure~\ref{fig:class} is shown in green, with solid line (scalp) and dashed line (cEEGrid). In general with scalp sensors, the classification accuracy was $1-3$ percentage points lower for \convnameshort compared to \sustnameshort (blue). The cEEGrid sensor classification, on the other hand, reached above 50\% accuracy for window lengths larger than \SI{8}{\second}, although this is not necessarily significant.

\subsubsection{Generalization of models}
\label{sec:res:crossclass}

To assess generalization of models trained on one condition, backward models fitted on \sustnameshort were applied to \switnameshort and \convnameshort data. Using a classification approach based on correlation between reconstructed and actual speech over \SI{35}{s}, mean accuracy was 0.70 (10th percentile $P_{10} = 0.47$, 90th percentile $P_{90}=0.89$) for \switnameshort and 0.66 ($P_{10} = 0.50$, $P_{90} = 0.78$) for \convnameshort. These results indicate that models trained on sustained attention generalize reasonably well to other attention conditions.

\section{Discussion}
\label{sec:dis}

\subsection{Key findings and contributions}

Our main finding is that selective attention can be reliably tracked across diverse listening conditions—from a simple sustained task to a dynamic, conversational setting—using a mobile EEG system with TRF-based methods. Unlike much of the existing literature, which relies on controlled audio-only stimuli, our approach incorporates real, uncontrolled AV speech, bringing neural tracking closer to everyday listening. Statistically significant differences in the P2 peak of the TRFs between attended and ignored speech (Figure~\ref{fig:stattrf}), widely associated with selective attention and is often reduced for ignored speech \cite{luck2012erp}, confirm that our mobile EEG setup effectively captures attention-related neural signatures in these complex conditions. We also observed significant clusters around the N400 peak, reflecting higher-order processing \cite{kutas1983event, ERPs}.

Our models proved stable and robust to attention switches and dynamic conversations, demonstrated by clear differences in reconstruction accuracy between attended and ignored speech (Figure~\ref{fig:recon}). This demonstrates the generalizability of our models across conditions, a capability previously unproven for mobile EEG in AV contexts. While mobile EEG has been validated for single-talker speech in various tasks \cite{Zink2016-eg,straetmans_neural_2024,Holle2021-vy}, its performance in switching tasks and conversational data, specifically with AV stimuli, was a significant gap, which this work addresses. We also showed that attention can be decoded in all three conditions using mobile EEG, with classification rates of 65-75\% for \SI{35}{\second} segments and significant rates for segments lasting a few seconds. Models were also effective when trained on one condition and applied to others, showing an invariance to task and stimuli type, supporting practical applications for real-world attention tracking.

\subsection{Broader implications and limitations}

\subsubsection{\textbf{Attention modulation across  listening conditions}}
The mobile EEG system captured consistent attention-related signatures across tasks. In \sustnameshort, the P2 peak confirmed the neural enhancement of attended signals in multi-talker environments \cite{luck2012erp}. In \switnameshort, attention tracking performances remained stable, showing that mobile EEG can follow dynamic reorientation of attention \cite{carta_simultaneous_2025}.

Conversational listening (\convnameshort) posed greater challenges. The TRF morphology was elongated with less distinct P1 and N1 peaks compared to the single-talker tasks. This likely reflects increased cognitive load when integrating multiple auditory and visual streams. This aligns with cognitive effort accounts, where resource allocation scales with task difficulty \cite{westbrook2015cognitive}. The conversational task required participants to track several sources of information simultaneously, resulting in higher load and less transient TRF responses, pointing to a shift toward more effortful and distributed processing.

\subsubsection{\textbf{Models generalize between tasks and AV speech material}}

Our forward and backward models demonstrated robust performance across multiple variables, showing strong generalizability. Firstly, the methodology was largely invariant to AV speech material. Attention tracking performance for conversation material (\convnameshort) was comparable to single talker speech (\sustnameshort and \switnameshort). This suggests that the models can capture attention-related neural signatures across a range of speech inputs. Second, we observed invariance to the listening task. Specifically, the attention switching in \switnameshort did not substantially affect the TRFs or classification performance, which remained comparable to the sustained attention task. Overall, these results demonstrate that a single model architecture with consistent hyperparameters can generalize across tasks and stimuli types, supporting the use of mobile EEG for real-world attention decoding. The stability across conditions also suggests that the neural features captured by our models are robust markers of selective attention rather than being tied to a specific task or speech stimulus.

\subsubsection{\textbf{Acoustics envelope vs. acoustic onsets}}
\label{sec:disc:features}
Several differences between the acoustic envelope and acoustic onset are evident from Figures \ref{fig:trf} and \ref{fig:stattrf}. Across all conditions, TRFs for individual electrodes were sharper and sparser for acoustic onsets than for acoustic envelopes, particularly for ignored speech (Figure~\ref{fig:trf} D, H and L), where the responses diminished after a dominant P1 peak around 60-\SI{80}{\milli\second}. This reflects the transient nature of acoustic onsets, which capture sudden amplitude increases, versus the continuous fluctuations captured by the acoustic envelope  \cite{rosenkranz}.

For \sustnameshort and \switnameshort, the dominant TRF peak for the acoustic envelope was P2, while for the acoustic onset it was P1. 
This aligns with the idea that the dense envelope tracks sustained stimulus dynamics, making it more sensitive to selective attention \cite{brodbeck_neural_2020}. Interestingly, this distinction was less pronounced in frontal-central electrodes during conversational tasks (\convnameshort, Figure~\ref{fig:trf}I), likely due to the higher attentional demands reported by participants (Figure~\ref{fig:behav}).

We observed a consistent latency difference of approximately \SI{20}{\milli\second} for P1, N1, and P2 between envelope and onset features across all conditions, consistent with prior TRF \cite{brodbeck_neural_2020}. This may result from the discrete nature of onsets, which elicit earlier peaks than the corresponding envelope signals. Alternatively, onsets may act as attentional cues, directing neural processing toward specific spectrotemporal regions and facilitating early envelope tracking \cite{lalor_2009, HauptENEURO.0287-24.2024, brodbeck_neural_2020}.

Aligned with prior studies \cite{brodbeck_eelbrain_2023}, the acoustic onsets elicited larger TRF peaks than the acoustic envelopes across all conditions. This likely reflects the well-established sensitivity of the auditory cortex to transient sound events \cite{DAUBE20191924}. Notably, this amplitude difference was particularly pronounced for ignored speech, suggesting that responses to sudden acoustic changes remain strong even when attention is directed elsewhere, in line with earlier observations \cite{brodbeck_neural_2020}.

\subsubsection{\textbf{cEEGrid performance and potential improvements}}
\label{sec:disc:ceegrid}
cEEGrid signals captured selective attention effects, but they were weaker compared to scalp EEG, limiting practical attention decoding. This aligns with previous findings, indicating that auditory decoding performance using cEEGrid currently lags behind scalp EEG \cite{mirkovic_target_2016, Nogueira2019-jk}. Improvements could include optimized electrode selection, noise removal, and referencing, potentially on a subject-specific basis given sufficient data \cite{holtze_ear-eeg_2022}. Additionally, incorporating simultaneous scalp recordings during model training could improve the accuracy of models relying solely on cEEGrid data.

\subsubsection{\textbf{TRFs for conversations vs. single talker}} 
\label{sec:disc:singleVSconv}
As shown in Figure~\ref{fig:conv}, TRFs derived from conversational material are more extended over time compared to TRFs for the single talker on the side, within the same conversation condition. In other words, the TRFs are less sparse and spread across longer time lags. This broadening likely arise from overlapping acoustic features or time-locked neural responses to the non-attended talker included in the fitting data. Additionally, the conversational stimuli included visual information, whereas the single talker on the side did not, which may further influence TRF shape. Behavioral studies also suggest that listening and speaking strategies adapt to the complexity of multi-speaker environments, further affecting neural responses  \cite{hadley2021conversation}. These differences should be considered in real-time neural tracking applications, where perfect separation of speech streams is often not achievable.

\section{Conclusion}
\label{sec:con}
Mobile EEG can reliably track selective attention in realistic listening scenarios beyond single-talker audio-only tasks. In this study, scalp EEG captured attention across three conditions: sustained attention to single-talker AV stimuli, attention switching between talkers, and attending to conversational AV sources. This was evidenced by clear P2 differences in TRFs between attended and ignored speech, as well as the performance of forward and backward models. The models remained robust across conditions, showing no significant performance drop during attention switches. Notably, TRF characteristics differed between single-talker and multi-talker conversations, which should be considered in future research. Attention modulation was weaker for cEEGrid data, highlighting the need for further methodological improvements to reliably track auditory attention.

Classification of selective attention was above chance across all conditions using mobile EEG scalp data, with accuracies ranging from $55\%-70\%$ for decision windows of $\SI{1.1}{\second}$ to $\SI{35}{\second}$. Backward models generalized across listening tasks and stimuli: models trained on sustained single-talker attention performed equally well on conversational and attention-switching tasks. These results advance the development of neural tracking systems for realistic auditory environments.

\section*{References}
\vspace{-2em}
\bibliographystyle{IEEEtran}
\bibliography{UnsupervisedAAD,MobileEEGpaper}

@article{lunner2020three,
  title={Three new outcome measures that tap into cognitive processes required for real-life communication},
  author={Lunner, Thomas and Alickovic, Emina and Graversen, Carina and Ng, Elaine Hoi Ning and Wendt, Dorothea and Keidser, Gitte},
  journal={Ear and hearing},
  volume={41},
  pages={39S--47S},
  year={2020},
  publisher={LWW}
}

@article{stapells2002cortical,
  title={Cortical event-related potentials to auditory stimuli},
  author={Stapells, David R},
  journal={Handbook of clinical audiology},
  volume={5},
  pages={378--406},
  year={2002},
  publisher={Williams \& Wilkins, Philadelphia, Pa, USA}
}

@article{o2015attentional,
  title={Attentional selection in a cocktail party environment can be decoded from single-trial \uppercase{EEG}},
  author={O'sullivan, James A and Power, Alan J and Mesgarani, Nima and Rajaram, Siddharth and Foxe, John J and Shinn-Cunningham, Barbara G and Slaney, Malcolm and Shamma, Shihab A and Lalor, Edmund C},
  journal={Cerebral cortex},
  volume={25},
  number={7},
  pages={1697--1706},
  year={2015},
  publisher={Oxford University Press}
}

@article{geirnaert2021electroencephalography,
  title={Electroencephalography-based auditory attention decoding: Toward neurosteered hearing devices},
  author={Geirnaert, Simon and Vandecappelle, Servaas and Alickovic, Emina and De Cheveigne, Alain and Lalor, Edmund and Meyer, Bernd T and Miran, Sina and Francart, Tom and Bertrand, Alexander},
  journal={IEEE Signal Processing Magazine},
  volume={38},
  number={4},
  pages={89--102},
  year={2021},
  publisher={IEEE}
}

@article{alickovic2019tutorial,
  title={A tutorial on auditory attention identification methods},
  author={Alickovic, Emina and Lunner, Thomas and Gustafsson, Fredrik and Ljung, Lennart},
  journal={Frontiers in neuroscience},
  volume={13},
  pages={153},
  year={2019},
  publisher={Frontiers Media SA}
}

@article{haro2022eeg,
  title={\uppercase{EEG} alpha and pupil diameter reflect endogenous auditory attention switching and listening effort},
  author={Haro, Stephanie and Rao, Hrishikesh M and Quatieri, Thomas F and Smalt, Christopher J},
  journal={European Journal of Neuroscience},
  volume={55},
  number={5},
  pages={1262--1277},
  year={2022},
  publisher={Wiley Online Library}
}

@article{peirce_psychopy2_2019,
	title = {{PsychoPy2}: {Experiments} in behavior made easy},
	volume = {51},
	issn = {1554-3528},
	shorttitle = {{PsychoPy2}},
	doi = {10.3758/s13428-018-01193-y},
	language = {en},
	number = {1},
	journal = {Behavior Research Methods},
	author = {Peirce, Jonathan and Gray, Jeremy R. and Simpson, Sol and MacAskill, Michael and Höchenberger, Richard and Sogo, Hiroyuki and Kastman, Erik and Lindeløv, Jonas Kristoffer},
	month = feb,
	year = {2019},
	pages = {195--203},
}

@misc{noauthor_sccnlabstreaminglayer_2025,
	title = {sccn/labstreaminglayer},
	url = {https://github.com/sccn/labstreaminglayer},
	urldate = {2025-06-18},
	publisher = {Swartz Center for Computational Neuroscience},
	month = jun,
	year = {2025},
	note = {original-date: 2018-02-28T10:50:12Z},
}

@article{straetmans_neural_2024,
	title = {Neural speech tracking and auditory attention decoding in everyday life},
	volume = {18},
	issn = {1662-5161},
	doi = {10.3389/fnhum.2024.1483024},
	journal = {Frontiers in Human Neuroscience},
	author = {Straetmans, Lisa and Adiloglu, Kamil and Debener, Stefan},
	month = nov,
	year = {2024},
	pages = {1483024},
}

@article{bleichner_identifying_2016,
	title = {Identifying auditory attention with ear-{EEG}: {cEEGrid} versus high-density cap-{EEG} comparison},
	volume = {13},
	issn = {1741-2560, 1741-2552},
	shorttitle = {Identifying auditory attention with ear-{EEG}},
	doi = {10.1088/1741-2560/13/6/066004},
	number = {6},
	journal = {Journal of Neural Engineering},
	author = {Bleichner, Martin G and Mirkovic, Bojana and Debener, Stefan},
	month = dec,
	year = {2016},
	pages = {066004},
}

@article{fishbach_auditory_2001,
	title = {Auditory {Edge} {Detection}: {A} {Neural} {Model} for {Physiological} and {Psychoacoustical} {Responses} to {Amplitude} {Transients}},
	volume = {85},
	issn = {0022-3077, 1522-1598},
	shorttitle = {Auditory {Edge} {Detection}},
	doi = {10.1152/jn.2001.85.6.2303},
	language = {en},
	number = {6},
	journal = {Journal of Neurophysiology},
	author = {Fishbach, Alon and Nelken, Israel and Yeshurun, Yehezkel},
	month = jun,
	year = {2001},
	pages = {2303--2323},
}

@article{brodbeck_neural_2020,
	title = {Neural speech restoration at the cocktail party: {Auditory} cortex recovers masked speech of both attended and ignored speakers},
	volume = {18},
	issn = {1545-7885},
	shorttitle = {Neural speech restoration at the cocktail party},
	doi = {10.1371/journal.pbio.3000883},
    language = {en},
	number = {10},
	journal = {PLOS Biology},
	author = {Brodbeck, Christian and Jiao, Alex and Hong, L. Elliot and Simon, Jonathan Z.},
	editor = {Malmierca, Manuel S.},
	month = oct,
	year = {2020},
	pages = {e3000883},
}

@article{brodbeck_eelbrain_2023,
	title = {Eelbrain, a {Python} toolkit for time-continuous analysis with temporal response functions},
	volume = {12},
	issn = {2050-084X},
	doi = {10.7554/eLife.85012},	
	language = {en},
	journal = {eLife},
	author = {Brodbeck, Christian and Das, Proloy and Gillis, Marlies and Kulasingham, Joshua P and Bhattasali, Shohini and Gaston, Phoebe and Resnik, Philip and Simon, Jonathan Z},
	month = nov,
	year = {2023},
	pages = {e85012},
}

@misc{di_liberto_speech_2025,
	title = {Speech {Neurophysiology} in {Realistic} {Contexts}: {Big} {Hype} or {Big} {Leap}?},
	copyright = {Creative Commons Attribution 4.0 International},
	shorttitle = {Speech {Neurophysiology} in {Realistic} {Contexts}},
	doi = {10.48550/ARXIV.2506.05494},
	publisher = {arXiv},
	author = {Di Liberto, Giovanni M. and Ip, Emily Y. J.},
	year = {2025},
	note = {Version Number: 1},
}

@article{zhang_boosting_2005,
	title = {Boosting with early stopping: {Convergence} and consistency},
	volume = {33},
	issn = {0090-5364},
	shorttitle = {Boosting with early stopping},
	doi = {10.1214/009053605000000255},
	number = {4},
	journal = {The Annals of Statistics},
	author = {Zhang, Tong and Yu, Bin},
	month = aug,
	year = {2005},
	note = {Publisher: Institute of Mathematical Statistics},
}

@article{david_estimating_2007,
	title = {Estimating sparse spectro-temporal receptive fields with natural stimuli},
	volume = {18},
	issn = {0954-898X, 1361-6536},
	doi = {10.1080/09548980701609235},
	language = {en},
	number = {3},
	journal = {Network: Computation in Neural Systems},
	author = {David, Stephen V. and Mesgarani, Nima and Shamma, Shihab A.},
	month = jan,
	year = {2007},
	note = {Publisher: Informa UK Limited},
	pages = {191--212},
}

@misc{van_de_ryck_eeg-based_2025,
	title = {{EEG}-based {Decoding} of {Auditory} {Attention} to {Conversations} with {Turn}-taking {Speakers}},
	copyright = {http://creativecommons.org/licenses/by/4.0/},
	url = {http://biorxiv.org/lookup/doi/10.1101/2025.06.20.660726},
	doi = {10.1101/2025.06.20.660726},
	urldate = {2025-07-16},
	publisher = {Cold Spring Harbor Laboratory},
	author = {Van De Ryck, Iris and Heintz, Nicolas and Rotaru, Iustina and Geirnaert, Simon and Bertrand, Alexander and Francart, Tom},
	month = jun,
	year = {2025},
}

@article{rotaru2024we,
  title={What are we really decoding? Unveiling biases in \uppercase{EEG}-based decoding of the spatial focus of auditory attention},
  author={Rotaru, Iustina and Geirnaert, Simon and Heintz, Nicolas and Van de Ryck, Iris and Bertrand, Alexander and Francart, Tom},
  journal={Journal of Neural Engineering},
  volume={21},
  number={1},
  pages={016017},
  year={2024},
  publisher={IOP Publishing}
}

@misc{carta_simultaneous_2025,
	title = {Simultaneous cortical tracking of competing speech streams during attention switching},
	copyright = {http://creativecommons.org/licenses/by/4.0/},
	url = {http://biorxiv.org/lookup/doi/10.1101/2025.07.02.662762},
	doi = {10.1101/2025.07.02.662762},
	urldate = {2025-07-16},
	publisher = {Cold Spring Harbor Laboratory},
	author = {Carta, Sara and Aličković, Emina and Zaar, Johannes and Valdés, Alejandro López and Di Liberto, Giovanni M.},
	month = jul,
	year = {2025},
}

@article{hadley2021conversation,
  title={Conversation in small groups: Speaking and listening strategies depend on the complexities of the environment and group},
  author={Hadley, Lauren V and Whitmer, William M and Brimijoin, W Owen and Naylor, Graham},
  journal={Psychonomic Bulletin \& Review},
  volume={28},
  number={2},
  pages={632--640},
  year={2021},
  publisher={Springer}
}

@article{westbrook2015cognitive,
  title={Cognitive effort: A neuroeconomic approach},
  author={Westbrook, Andrew and Braver, Todd S},
  journal={Cognitive, Affective, \& Behavioral Neuroscience},
  volume={15},
  number={2},
  pages={395--415},
  year={2015},
  publisher={Springer}
}

@article{kutas1983event,
  title={Event-related brain potentials to grammatical errors and semantic anomalies},
  author={Kutas, Marta and Hillyard, Steven A},
  journal={Memory \& cognition},
  volume={11},
  number={5},
  pages={539--550},
  year={1983},
  publisher={Springer}
}

@article{holtze_ear-eeg_2022,
	title = {Ear-{EEG} {Measures} of {Auditory} {Attention} to {Continuous} {Speech}},
	volume = {16},
	copyright = {https://creativecommons.org/licenses/by/4.0/},
	issn = {1662-453X},
	doi = {10.3389/fnins.2022.869426},
	journal = {Frontiers in Neuroscience},
	author = {Holtze, Björn and Rosenkranz, Marc and Jaeger, Manuela and Debener, Stefan and Mirkovic, Bojana},
	month = may,
	year = {2022},
	note = {Publisher: Frontiers Media SA},
}

@article{mirkovic_target_2016,
	title = {Target {Speaker} {Detection} with {Concealed} {EEG} {Around} the {Ear}},
	volume = {10},
	issn = {1662-453X},
	doi = {10.3389/fnins.2016.00349},
	journal = {Frontiers in Neuroscience},
	author = {Mirkovic, Bojana and Bleichner, Martin G. and De Vos, Maarten and Debener, Stefan},
	month = jul,
	year = {2016},
	note = {Publisher: Frontiers Media SA},
}

@article{debener2015unobtrusive,
  title={Unobtrusive ambulatory {EEG} using a smartphone and flexible printed electrodes around the ear},
  author={Debener, Stefan and Emkes, Reiner and De Vos, Maarten and Bleichner, Martin},
  journal={Scientific reports},
  volume={5},
  number={1},
  pages={16743},
  year={2015},
  publisher={Nature Publishing Group UK London}
}

@article{kappel2018dry,
  title={Dry-contact electrode ear-{EEG}},
  author={Kappel, Simon L and Rank, Mike L and Toft, Hans Olaf and Andersen, Mikael and Kidmose, Preben},
  journal={IEEE Transactions on Biomedical Engineering},
  volume={66},
  number={1},
  pages={150--158},
  year={2018},
  publisher={IEEE}
}

@article{keidser2020quest,
  title={The quest for ecological validity in hearing science: What it is, why it matters, and how to advance it},
  author={Keidser, Gitte and Naylor, Graham and Brungart, Douglas S and Caduff, Andreas and Campos, Jennifer and Carlile, Simon and Carpenter, Mark G and Grimm, Giso and Hohmann, Volker and Holube, Inga and others},
  journal={Ear and hearing},
  volume={41},
  pages={5S--19S},
  year={2020},
  publisher={LWW}
}

@article{bodie2023listening,
  title={Listening as a positive communication process},
  author={Bodie, Graham D},
  journal={Current Opinion in Psychology},
  volume={53},
  pages={101681},
  year={2023},
  publisher={Elsevier}
}

@article{belo2021eeg,
  title={\uppercase{EEG}-based auditory attention detection and its possible future applications for passive \uppercase{BCI}},
  author={Belo, Joan and Clerc, Maureen and Sch{\"o}n, Daniele},
  journal={Frontiers in computer science},
  volume={3},
  pages={661178},
  year={2021},
  publisher={Frontiers Media SA}
}

@article{schafer2018testing,
  title={Testing the limits of the stimulus reconstruction approach: auditory attention decoding in a four-speaker free field environment},
  author={Sch{\"a}fer, Patrick J and Corona-Strauss, Farah I and Hannemann, Ronny and Hillyard, Steven A and Strauss, Daniel J},
  journal={Trends in hearing},
  volume={22},
  pages={2331216518816600},
  year={2018},
  publisher={SAGE Publications Sage CA: Los Angeles, CA}
}

@article{ciccarelli2019comparison,
  title={Comparison of two-talker attention decoding from \uppercase{EEG} with nonlinear neural networks and linear methods},
  author={Ciccarelli, Gregory and Nolan, Michael and Perricone, Joseph and Calamia, Paul T and Haro, Stephanie and O’sullivan, James and Mesgarani, Nima and Quatieri, Thomas F and Smalt, Christopher J},
  journal={Scientific reports},
  volume={9},
  number={1},
  pages={11538},
  year={2019},
  publisher={Nature Publishing Group UK London}
}

@article{tanveer2024deep,
  title={Deep learning-based auditory attention decoding in listeners with hearing impairment},
  author={Tanveer, M Asjid and Skoglund, Martin A and Bernhardsson, Bo and Alickovic, Emina},
  journal={Journal of Neural Engineering},
  volume={21},
  number={3},
  pages={036022},
  year={2024},
  publisher={IOP Publishing}
}

@article{alickovic2021effects,
  title={Effects of hearing aid noise reduction on early and late cortical representations of competing talkers in noise},
  author={Alickovic, Emina and Ng, Elaine Hoi Ning and Fiedler, Lorenz and Santurette, S{\'e}bastien and Innes-Brown, Hamish and Graversen, Carina},
  journal={Frontiers in neuroscience},
  volume={15},
  pages={636060},
  year={2021},
  publisher={Frontiers Media SA}
}

@misc {Kriegstein2021,
	Title = {A Multisensory Perspective on Human Auditory Communication},
	Author = {von Kriegstein, Katharina},
	Publisher = {CRC Press/Taylor &amp; Francis, Boca Raton (FL)},
	Year = {2012},
	ISBN = {9781439812174},
	Series = {Frontiers in Neuroscience},
	URL = {http://europepmc.org/books/NBK92846},
}

@article{wang2023eeg,
  title={\uppercase{EEG}-based auditory attention decoding with audiovisual speech for hearing-impaired listeners},
  author={Wang, Bo and Xu, Xiran and Niu, Yadong and Wu, Chao and Wu, Xihong and Chen, Jing},
  journal={Cerebral Cortex},
  volume={33},
  number={22},
  pages={10972--10983},
  year={2023},
  publisher={Oxford University Press}
}

@article{fu2019congruent,
  title={Congruent audiovisual speech enhances auditory attention decoding with \uppercase{EEG}},
  author={Fu, Zhen and Wu, Xihong and Chen, Jing},
  journal={Journal of neural engineering},
  volume={16},
  number={6},
  pages={066033},
  year={2019},
  publisher={IOP Publishing}
}

@article{puffay2023relating,
  title={Relating \uppercase{EEG} to continuous speech using deep neural networks: a review},
  author={Puffay, Corentin and Accou, Bernd and Bollens, Lies and Monesi, Mohammad Jalilpour and Vanthornhout, Jonas and Francart, Tom and others},
  journal={Journal of Neural Engineering},
  volume={20},
  number={4},
  pages={041003},
  year={2023},
  publisher={IOP Publishing}
}

@article{lee1999independent,
  title={Independent component analysis using an extended infomax algorithm for mixed subgaussian and supergaussian sources},
  author={Lee, Te-Won and Girolami, Mark and Sejnowski, Terrence J},
  journal={Neural computation},
  volume={11},
  number={2},
  pages={417--441},
  year={1999},
  publisher={MIT Press}
}

@article{luck2012erp,
  title={\uppercase{ERP} components and selective attention},
  author={Luck, Steven J and Kappenman, Emily S},
  journal={The Oxford handbook of event-related potential components},
  pages={295--327},
  year={2012},
  publisher={Oxford University Press New York}
}

@article{o2019look,
  title={Look at me when I'm talking to you: Selective attention at a multisensory cocktail party can be decoded using stimulus reconstruction and alpha power modulations},
  author={O'Sullivan, Aisling E and Lim, Chantelle Y and Lalor, Edmund C},
  journal={European Journal of Neuroscience},
  volume={50},
  number={8},
  pages={3282--3295},
  year={2019},
  publisher={Wiley Online Library}
}

@article{crosse2015congruent,
  title={Congruent visual speech enhances cortical entrainment to continuous auditory speech in noise-free conditions},
  author={Crosse, Michael J and Butler, John S and Lalor, Edmund C},
  journal={Journal of Neuroscience},
  volume={35},
  number={42},
  pages={14195--14204},
  year={2015},
  publisher={Society for Neuroscience}
}

@ARTICLE{ERPs,
  
AUTHOR={Proverbio, Alice Mado  and Tacchini, Marta  and Jiang, Kaijun },
         
TITLE={Event-related brain potential markers of visual and auditory perception: A useful tool for brain computer interface systems},
        
JOURNAL={Frontiers in Behavioral Neuroscience},
        
VOLUME={Volume 16 - 2022},

YEAR={2022},

DOI={10.3389/fnbeh.2022.1025870},

ISSN={1662-5153}}

@ARTICLE{Smith2009-fw,
  title     = "Threshold-free cluster enhancement: addressing problems of
               smoothing, threshold dependence and localisation in cluster
               inference",
  author    = "Smith, Stephen M and Nichols, Thomas E",
  journal   = "Neuroimage",
  publisher = "Elsevier BV",
  volume    =  44,
  number    =  1,
  pages     = "83--98",
  month     =  jan,
  year      =  2009,
  language  = "en"
}

@article{workingmemory1,
author = {Salisbury, Dean},
year = {2004},
month = {08},
pages = {396-9},
title = {Semantic memory and verbal working memory correlates of N400 to subordinate homographs},
volume = {55},
journal = {Brain and cognition},
doi = {10.1016/j.bandc.2004.02.057}
}

@article{workingmemory2,
author = {Kutas, Marta and Federmeier, Kara},
year = {2011},
month = {01},
pages = {621-47},
title = {{Thirty Years and Counting: Finding Meaning in the N400 Component of the Event-Related Brain Potential (ERP)}},
volume = {62},
journal = {Annual review of psychology},
doi = {10.1146/annurev.psych.093008.131123}
}

@article {GillisENEURO.0075-23.2023,
	author = {Gillis, Marlies and Vanthornhout, Jonas and Francart, Tom},
	title = {Heard or Understood? Neural Tracking of Language Features in a Comprehensible Story, an Incomprehensible Story and a Word List},
	volume = {10},
	number = {7},
	elocation-id = {ENEURO.0075-23.2023},
	year = {2023},
	doi = {10.1523/ENEURO.0075-23.2023},
	publisher = {Society for Neuroscience},
	eprint = {https://www.eneuro.org/content/10/7/ENEURO.0075-23.2023.full.pdf},
	journal = {eNeuro}
}

@ARTICLE{Nogueira2019-jk,
  title     = "Decoding selective attention in normal hearing listeners and
               bilateral cochlear implant users with concealed ear {EEG}",
  author    = "Nogueira, Waldo and Dolhopiatenko, Hanna and Schierholz, Irina
               and B{\"u}chner, Andreas and Mirkovic, Bojana and Bleichner,
               Martin G and Debener, Stefan",
  journal   = "Front. Neurosci.",
  publisher = "Frontiers Media SA",
  volume    =  13,
  pages     = "720",
  month     =  jul,
  year      =  2019,
  copyright = "https://creativecommons.org/licenses/by/4.0/",
  language  = "en"
}

@unknown{rosenkranz,
author = {Rosenkranz, Marc and Haupt, Thorge and Jaeger, Manuela and Uslar, Verena Nicole and Bleichner, Martin},
year = {2024},
month = {05},
pages = {},
title = {{Sound perception in realistic surgery scenarios: Towards EEG-based auditory work strain measures for medical personnel}},
doi = {10.1101/2024.05.07.592873}
}

@ARTICLE{Holle2021-vy,
  title     = "Mobile {ear-EEG} to study auditory attention in everyday life :
               Auditory attention in everyday life",
  author    = "H{\"o}lle, Daniel and Meekes, Joost and Bleichner, Martin G",
  journal   = "Behav. Res. Methods",
  publisher = "Springer Science and Business Media LLC",
  volume    =  53,
  number    =  5,
  pages     = "2025--2036",
  month     =  oct,
  year      =  2021,
  copyright = "https://creativecommons.org/licenses/by/4.0",
  language  = "en"
}

@ARTICLE{Zink2016-eg,
  title     = "Mobile {EEG} on the bike: disentangling attentional and physical
               contributions to auditory attention tasks",
  author    = "Zink, Rob and Hunyadi, Borb{\'a}la and Van Huffel, Sabine and
               Vos, Maarten De",
  journal   = "J. Neural Eng.",
  publisher = "IOP Publishing",
  volume    =  13,
  number    =  4,
  pages     = "046017",
  month     =  aug,
  year      =  2016,
  copyright = "http://iopscience.iop.org/page/copyright"
}

@article{lalor_2009,
author = {Lalor, Edmund C. and Power, Alan J. and Reilly, Richard B. and Foxe, John J.},
title = {{Resolving Precise Temporal Processing Properties of the Auditory System Using Continuous Stimuli}},
journal = {Journal of Neurophysiology},
volume = {102},
number = {1},
pages = {349-359},
year = {2009},
doi = {10.1152/jn.90896.2008},
    note ={PMID: 19439675},
eprint = { https://doi.org/10.1152/jn.90896.2008}
}

@article {HauptENEURO.0287-24.2024,
	author = {Haupt, Thorge and Rosenkranz, Marc and Bleichner, Martin G.},
	title = {{Exploring Relevant Features for EEG-Based Investigation of Sound Perception in Naturalistic Soundscapes}},
	volume = {12},
	number = {1},
	elocation-id = {ENEURO.0287-24.2024},
	year = {2025},
	doi = {10.1523/ENEURO.0287-24.2024},
	publisher = {Society for Neuroscience},
	eprint = {https://www.eneuro.org/content/12/1/ENEURO.0287-24.2024.full.pdf},
	journal = {eNeuro}
}

@article{DAUBE20191924,
title = {Simple Acoustic Features Can Explain Phoneme-Based Predictions of Cortical Responses to Speech},
journal = {Current Biology},
volume = {29},
number = {12},
pages = {1924-1937.e9},
year = {2019},
issn = {0960-9822},
doi = {https://doi.org/10.1016/j.cub.2019.04.067},
author = {Christoph Daube and Robin A.A. Ince and Joachim Gross}
}

@misc{tobii,
	title = {Tobii Pro Nano},
	url = {https://www.tobii.com/products/discontinued/tobii-pro-nano},
	urldate = {2025-09-16},
	publisher = {Tobii},
	year = {2025},
}

@article{benjaminihochberg,
 ISSN = {00359246},
 author = {Yoav Benjamini and Yosef Hochberg},
 journal = {Journal of the Royal Statistical Society. Series B (Methodological)},
 number = {1},
 pages = {289--300},
 publisher = {[Royal Statistical Society, Oxford University Press]},
 title = {Controlling the False Discovery Rate: A Practical and Powerful Approach to Multiple Testing},
 volume = {57},
 year = {1995}
}

@article{geirnaert_time-adaptive_2022,
	title = {Time-{Adaptive} {Unsupervised} {Auditory} {Attention} {Decoding} {Using} {EEG}-{Based} {Stimulus} {Reconstruction}},
	volume = {26},
	copyright = {https://ieeexplore.ieee.org/Xplorehelp/downloads/license-information/IEEE.html},
	issn = {2168-2194, 2168-2208},
	url = {https://ieeexplore.ieee.org/document/9743715/},
	doi = {10.1109/JBHI.2022.3162760},
	number = {8},
	urldate = {2025-02-21},
	journal = {IEEE Journal of Biomedical and Health Informatics},
	author = {Geirnaert, Simon and Francart, Tom and Bertrand, Alexander},
	month = aug,
	year = {2022},
	pages = {3767--3778},
}

\end{document}